\documentclass[12pt]{article}
\pdfoutput=1
\usepackage{jhep-mod}
\usepackage{bm}
\usepackage{amssymb, amsmath}
\usepackage{pifont}
\usepackage{slashed} 
\usepackage{slantsc} 
\usepackage{hyperref}
\usepackage{graphicx}
\usepackage{datetime}
\def\d{{\mathrm{d}}}

\def\O{{\mathcal{O}}}
\def\I{{\mathcal{I}}}
\def\Im{{\mathrm{Im}}}

\def\lint{\hbox{\Large $\displaystyle\int$}}   
\makeatletter
\graphicspath{{.}{./Figs/}{./pdf/}}

\title{
\centerline{Greybody factors for Schwarzschild black holes:}
\centerline{Path-ordered exponentials \& product integrals}
}
\author{Finnian Gray {\sf and}}
\emailAdd{finnian.gray@msor.vuw.ac.nz}
\author{Matt Visser\,}
\emailAdd{matt.visser@msor.vuw.ac.nz}
\affiliation{School of Mathematics and Statistics,
Victoria University of Wellington; \\
PO Box 600, Wellington 6140, New Zealand.}
\abstract{\\
In recent work concerning the sparsity of the Hawking flux [arXiv:1506.03975v2], we found it necessary to re-examine what is known regarding the greybody factors of black holes, with a view to extending and expanding on some old results from the 1970s. \\ 
Focussing specifically on Schwarzschild  black holes,  we re-calculated and re-assessed the greybody factors using a \emph{path-ordered-exponential} approach, a technique which has the virtue of providing a semi-explicit formula for the relevant Bogoliubov coefficients. These path-ordered-exponentials, (being based on a ``\emph{transfer matrix}'' formalism), are closely related to so-called ``\emph{product integrals}'', leading to quite straightforward and direct numerical evaluation, while avoiding any need for numerically solving differential equations. Furthermore, while considerable analytic information is already available regarding both the high-frequency and low-frequency asymptotics of these greybody factors, numerical approaches seem better adapted to finding suitable ``global models'' for these greybody factors in the intermediate frequency regime, where most of the Hawking flux is concentrated. 

Working in a more general context, these \emph{path-ordered-exponential} techniques are also likely to be of interest for generic barrier-penetration problems.

\medskip
\noindent
D{\sc{ate}}:  16 December 2015; \LaTeX-ed \today --- \currenttime
}
\enlargethispage{15pt}
\begin{document}
\maketitle
\clearpage
\section{Introduction}

In recent work~\cite{Gray:2015} on the sparsity of the Hawking flux the present authors, (and two others), found it necessary to numerically evaluate certain greybody factors for the transmission of massless bosons through the Regge--Wheeler potential --- the combined gravitational and angular momentum potential barrier surrounding a Schwarzschild black hole. While significant work on this topic dates back to the mid-1970's, see in particular references~\cite{Page:1976a, Page:1976b, Page:1977, Page:thesis}, we felt it useful to completely re-assess the situation in terms of a radically different formalism using \emph{path-ordered matrix exponentials}~\cite{Visser:1999, Boonserm:2010, Boonserm:2013}, a formalism closely related both to ``\emph{transfer matrices}'' and to the so-called ``\emph{product integral}"~\cite{Helton:1975,  Helton:1976, Dollard:1979, Dollard:1984,  Slavik:2007}. One of the virtues of using path-ordered matrix exponentials is that it gives a semi-explicit expression for the Bogoliubov coefficients, and so gives a somewhat deeper analytic understanding of the  transmission and reflection probabilities~\cite{Boonserm:2008a, Boonserm:2008c, Boonserm:2009}. Path-ordered matrix exponentials are also relatively simple to evaluate numerically, and one never actually has to solve a numerical differential equation,  one just performs a numerical integral. 

In counterpoint, there has also been a lot of work done on both low-frequency and high-frequency limits for the greybody factors. (See for instance references~\cite{Page:1976a, Starobinsky} for low-frequency  limits, and references~\cite{Sanchez:1976, Sanchez:1977a,Sanchez:1977b} for high-frequency limits.)  Information at intermediate frequencies is harder to come by, and we shall use our numerical insights to try to develop some semi-analytic understanding of the intermediate frequency regime. 

We emphasize that while in the current article we are interested in black hole physics, the 
\emph{path-ordered matrix exponential} formalism~\cite{Visser:1999, Boonserm:2010, Boonserm:2013} is a general-purpose tool, of wide applicability to both barrier penetration and scattering processes. 

\section{Strategy}

Consider a Schr\"odinger-like scattering problem of the form
\begin{equation}
\label{E:strategy}
- {\d^2\over \d x^2} \psi(x) + V(x) \; \psi(x) = \omega^2 \psi(x); 
\qquad\qquad
V(x\to\pm\infty) \to 0.
\end{equation}
For any such problem one can define Bogoliubov coefficients relating the asymptotic left and right free-particle states, and derive the exact path-ordered-exponential result~\cite{Visser:1999}:
\begin{equation}
\label{E:Transfer0}
\begin{bmatrix}
\alpha & \beta^* \\
\beta & \alpha^* \\
\end{bmatrix}
=
\mathcal{P}\exp\left(-\frac{i}{2\omega} \int_{-\infty}^{+\infty} V(x)
\begin{bmatrix}
1 & e^{-2i\omega x}\\
-e^{2i\omega x} & -1 \\
\end{bmatrix}
\d x \right).
\end{equation}
A brief sketch of the derivation is given in appendix \ref{A:A}; see also reference~\cite{Visser:1999} for details. This result, and various generalizations and modifications thereof, has been used in a number of subsequent articles to develop general bounds on transmission probabilities, both generic and black hole specific. See for instance references~\cite{Visser:1999, Boonserm:2010, Boonserm:2013}, additional formal developments in references~\cite{Boonserm:2008a, Boonserm:2008c, Boonserm:2009}, and some applications in references~\cite{Boonserm:2008b, Boonserm:2014a, Boonserm:2014b}. The key point is that the Bogoliubov coefficients satisfy $|\alpha|^2-|\beta|^2=1$, and that the transmission probability is simply
\begin{equation}
\label{E:Tprob}
T(\omega) = {1\over \alpha(\omega) \; \alpha^*(\omega)} =  {1\over |\alpha(\omega)|^2} . 
\end{equation}
In the current article, instead of looking for bounds, we shall use this technique as a basis for numerically calculating greybody factors for the Schwarzschild black hole. The key steps are to replace the position $x$ by the tortoise coordinate $r_*$, and to replace $V(x)$ by the Regge--Wheeler potential~\cite{Boonserm:2013, Regge:1957}. Some formal developments are relegated to appendix \ref{A:B}.

\section{Path-ordered-exponentials and the product calculus}
The definition of the path ordered exponential, for a real or complex valued $m\times m$ matrix function $A(x)$ from some initial point $x_i$ to some final point $x_f$, is simply:
\begin{equation}
\label{E:POE}
\mathcal{P}\exp\left(\int_{x_i}^{x_f}A(x)\;\d x\right)\equiv 
\lim_{N\rightarrow\infty} \prod_{k=0}^{N-1}\exp\big(A(x^*_k)\Delta x_k\big).
\end{equation}
Here, (closely following the usual technical construction of the Riemann integral),  we shall take $x^*_k\in[x_k,x_{k+1}]$ to be a tag of some partition $\{x_k\}_{k=0}^N$ of the interval $[x_i,x_f] = [x_0,x_N]$, with individual widths $\Delta x_k=(x_{k+1}-x_k)$, and with the mesh ($D=\max\left\{\Delta x_k\right\}$) going to zero in the limit $N\rightarrow\infty$. 

This definition of the path ordered exponential is equivalent to (a matrix-valued, non-commutative) definition of the so-called ``product integral''~\cite{ Helton:1975, Helton:1976, Dollard:1979,  Dollard:1984, Slavik:2007}. The product calculus is exactly the idea of defining a different notion of calculus based on infinitesimal products and divisions, (recall that ordinary calculus is based on infinitesimal sums and subtractions). In the language of the product calculus equation (\ref{E:POE}) becomes
\begin{equation}\label{E:ProdInt}
\mathcal{P}\exp\left(\int_{x_i}^{x_f}A(x) \;\d x\right)=\prod_{x_i}^{x_f}\left(\I +A(x) \;\d x\right).
\end{equation}
Here the product integral is defined as~\cite{Helton:1975,Helton:1976, Dollard:1979,  Dollard:1984, Slavik:2007}
\begin{equation}
\label{E:Prod}
\prod_{x_i}^{x_f}\left(\I +A(x) \;\d x\right)\equiv\lim_{N\rightarrow\infty} \prod_{k=0}^{N-1}\left(\I +A(x^*_k)\Delta x_k\right),
\end{equation}
with the partition $\{x_k\}_{k=0}^N$ of $[x_i,x_f]=[x_0,x_N]$, the tag $x^*_k$, the width $\Delta x_k$, and the mesh $D$, again being defined as before. The equivalence of the two definitions given in equations  (\ref{E:POE}) and (\ref{E:Prod}) can be deduced by noting that all the second order or higher times in $\exp(A(x^*_k)\Delta x_k)$ go to zero in the limit $D\rightarrow 0$~\cite{Dollard:1984}. (Note that if the matrix $A(x)$ happens to be nilpotent, so that $A(x)^2=0$, (which is quite often the case), then we have the \emph{exact} result that $\exp\big(A(x^*_k)\Delta x_k\big)= \I  + \big(A(x^*_k)\,\Delta x_k\big) $, so the equality holds even before the limit of vanishing mesh is taken.) See for instance references~\cite{Dollard:1984, Slavik:2007} for an overview of the product calculus. 

The following results for the product integral~\cite{Helton:1975,Helton:1976, Dollard:1979,  Dollard:1984, Slavik:2007} should be especially noted:
\begin{itemize}
\item 
If \;$\prod_a^b(\I +A \;\d x)$ exists, then $A(x)$ is bounded on $[a,b]$.

\item 
If \;$\prod_a^b(\I +A \;\d x)$ exists, and $a\leq u<v\leq b$, then $\prod_u^v(\I+A \;\d x)$ exists.

\item 
If \;$a<b<c$, and both $\prod_a^b(\I +A \;\d x)$ and $\prod_b^c(\I +A \;\d x)$ exist, then
\begin{equation}
\prod_a^c(\I +A \;\d x)=\prod_a^b(\I +A \;\d x)\times\prod_b^c(\I +A \;\d x).
\end{equation} 

\item 
$\prod_a^b(\I +A \;\d x)$ exists if and only if $\int_a^bA \;\d x$ exists.

\item 
If $\prod_a^b(\I +A \;\d x)$ exists, then
\begin{eqnarray}
\label{E:Peano}
\prod_a^b(\I +A \;\d x) 
&=&\I  +\int_a^b A(x_1) \;\d x_1+\int_a^b\int_a^{x_1}A(x_1)A(x_2) \;\d x_2 \;\d x_1 
\nonumber\\
&& + \int_a^b\int_a^{x_1}\int_a^{x_2}A(x_1)A(x_2)A(x_3) \;\d x_3 \;\d x_2 \;\d x_1 +...
\end{eqnarray}
\end{itemize}

Equation (\ref{E:Peano}) is commonly known in the mathematics community as the Peano series,  and in the physics community as the Dyson series. (There are also alternative expansions available in terms of the Magnus series and/or Fer series that might sometimes be useful. We shall not explore such options here.)  This Dyson series is the physics community's standard method for approximating path ordered exponentials. Using the path-ordering operator $\mathcal{P}$, equation (\ref{E:Peano}) can be re-written as
\begin{equation}
\prod_a^b(\I +A \;\d x)=\sum_{n=0}^{\infty}\frac{1}{n!}\int_{a}^{b} \;\d x_1\int_{a}^{b} \;\d x_2 \cdots\int_{a}^{b} \;\d x_n\; \mathcal{P}\left\{A(x_1)A(x_2)\cdots A(x_n)\right\},
\end{equation}
where now
\begin{equation}
\label{E:Peano2}
\mathcal{P}\left\{A(x_1)A(x_2)\cdots A(x_n)\right\}=A(x_{\sigma(1)})A(x_{\sigma(2)})\cdots A(x_{\sigma(n)}),
\end{equation}
with $\sigma_{(i)}$ being any permutation such that $x_{\sigma(1)}\geq x_{\sigma(2)}\geq...\geq x_{\sigma(n)}$. 
For example
\begin{equation}
\mathcal{P}\left\{A(x_1)A(x_2)\right\}=\Theta(x_1-x_2)A(x_1)A(x_2)+\Theta(x_2-x_1)A(x_2)A(x_1),
\end{equation}
where $\Theta(x)$ is the Heaviside step function.

When written in the form of equation (\ref{E:ProdInt}) the transfer matrix of equation (\ref{E:Transfer0}) is particularly simple to numerically evaluate. Defining
 \begin{equation}
A(x_k)\equiv
-\frac{i}{2\omega}V(x_k)
\begin{bmatrix}
1 & e^{-2i\omega x_k}\\
-e^{2i\omega x_k} & -1 \\
\end{bmatrix}\;,
\end{equation}
we note $A(x_k)^2=0$.
The transfer matrix can be compactly written as
 \begin{eqnarray}
E(x_i,x_f)&=&\lim_{N\rightarrow\infty} \prod_{k=1}^{N}\Big(\mathcal{I}+A(x_k)h \Big) \nonumber\\
		&=& \lim_{N\rightarrow\infty}\left\{\Big(\mathcal{I}+A(x_{N-1})h \Big)\cdots\Big(\mathcal{I}+A(x_2)h\Big)\Big(\mathcal{I}+A(x_1)h \Big)\right\} \;.
\end{eqnarray}
Here $h=(x_f-x_i)/N$, and  we have chosen the right-tagged equipartition $x_k=x_i+kh$. This naive expression is extremely easy to compute numerically; however much like a simple Riemann sum, convergence can sometimes be rather slow. In order to improve convergence, Helton and Stuckwisch~\cite{Helton:1976} introduce a polynomial approximation which has error bounded by $H(x_i,x_f)\,h^p$, where, $H(x_i,x_f)$ is a bounded interval function determined by the matrix $A$, and $p$ is the order of the polynomial approximation. 
(This is the product integral equivalent of a higher-order Simpson rule.)

A  minor technical issue is that Helton and Stuckwisch use a definition of the product integral based on right hand multiplication, that is, the order of the products in equation (\ref{E:Prod}) is reversed. This can be related to the form in equation (\ref{E:Prod}) by noting that in the limit $N\rightarrow\infty$,
\begin{align}
\left[\prod_{x_i}^{x_f}\left(\mathcal{I}+A(x)\;dx\right)\right]^{-1}=&
\left[\lim_{N\rightarrow\infty}\left\{\Big(\mathcal{I}+A(x_{N-1})h \Big)...\Big(\mathcal{I}+A(x_2)h\Big)\Big(\mathcal{I}+A(x_1)h \Big)\right\}\right]^{-1}\nonumber\\
=&\lim_{N\rightarrow\infty}\left[\Big(\mathcal{I}+A(x_{N-1})h \Big)...\Big(\mathcal{I}+A(x_2)h\Big)\Big(\mathcal{I}+A(x_1)h \Big)\right]^{-1}\nonumber\\
=&\lim_{N\rightarrow\infty}\left\{\Big(\mathcal{I}-A(x_{1})h \Big)\Big(\mathcal{I}-A(x_2)h\Big)...\Big(\mathcal{I}-A(x_{N-1})h \Big) \right\}\;.
\end{align}
Thus by sending $A(x_i)\rightarrow-A(x_i)$ and using the approximation in reference~\cite{Helton:1976} we obtain the inverse of the product integral we set out to calculate. Since the Bogoliubov matrix satisfies
\begin{equation}
\begin{bmatrix}
\alpha & \beta^*\\
\beta & \alpha^* \\
\end{bmatrix}^{-1}
=
\begin{bmatrix}
\alpha^* & -\beta^*\\
-\beta & \alpha \\
\end{bmatrix}\;,
\end{equation}
we see that there is no loss of information; this method lends itself very well to finding the transmission probabilities.

The 5th-order approximation Helton and Stuckwisch introduce is this~\cite{Helton:1976}:
\begin{align}
\label{E:Approx}
\prod_{k=1}^N&\left[ \mathcal{I}+\left(h/90\right) \left(7A_0+32A_1+12A_2+32A_3+7A_4\right) \right. \nonumber\\
&+(h^2/90)\Big(8A_0A_1-12A_0A_2+18A_0A_3-7A_0A_4+18A_1A_2-12A_1A_3 \nonumber\\
&\qquad\qquad+18A_1A_4+18A_2A_3-12A_2A_4+8A_3A_4\Big)\nonumber\\
&+(h^3/60)\Big(3A_0A_1A_2-2A_0A_1A_3+3A_0A_1A_4-2A_0A_2A_3-2A_0A_2A_4\nonumber\\
&\qquad\qquad+3A_0A_3A_4+8A_1A_2A_3-2A_1A_2A_4-2A_1A_3A_4+3A_2A_3A_4\Big)\nonumber\\
&+(h^4/120)\Big(4A_0A_1A_2A_3-A_0A_1A_2A_4-A_0A_1A_3A_4-A_0A_2A_3A_4+4a_1A_2A_3a_4\Big)\nonumber\\
&\left.+(h^5/120)A_0A_1A_2A_3A_4A_5\right]\;,
\end{align}
where $A_j=A[x_{k-1}+jh/4]$ for $j=0,1,2,3,4$ and $k=1,2,..,N$. This somewhat unwieldy expression can more compactly be rewritten as~\cite{Helton:1976}:
\begin{equation}
\label{E:Approx2}
\prod_{k=1}^N\left[\mathcal{I}+(28K_1+32K_2+6K_3+4K_4+K_5)/360\right]\;.
\end{equation}
Here
\begin{eqnarray*}
K_1&=&hA_4\\
K_2&=&hA_3(4\mathcal{I}+K_1)\\
K_3&=&hA_2[8(\mathcal{I}-K_1)+3K_2]\\
K_4&=&hA_1[32\mathcal{I}+18K_1+3hA_2(6\mathcal{I}-K_1+K_2)-3K_2]\\
K_5&=&hA_0[28(\mathcal{I}-K_1)-3hA_2(16\mathcal{I}+4K_1+K_2)+K_4+18K_2]\;.
\end{eqnarray*}
This scheme will be our primary tool for actually evaluating greybody factors.
We first tested our technique in two situations, (the square-barrier potential, the delta-function potential), where exact analytic results are known. We thereby verified that the numerical integration adequately approximated the exact results, (details mercifully suppressed), before turning to the Schwarzschild geometry and its associated Regge--Wheeler potential.

\section{Particle emission from a Schwarzschild black hole}

We now turn to the Regge--Wheeler potential, which is related to particle emission from black holes. In this situation there is no exact analytic expression for the transmission probability, neither by directly analysing incoming and outgoing waves, nor via the product calculus method. Instead we must evaluate the Bogoliubov coefficients numerically. 
(Oddly enough for this problem the \emph{exact} wavefunctions are known in terms of \emph{Heun functions}, though this observation is less useful than it might seem simply because not enough is understood regarding the asymptotic behaviour of these Heun functions. See references~\cite{Hortacsu:2011, Fiziev:2011,  Fiziev:2007, Fiziev:2006}.)
We shall work in ``geometric units'' where both $G_\mathrm{Newton}\to1$ and $c\to 1$.

\subsection{Setup}

The probability of emission, (the greybody factors), of massless particles from a black hole can be found by analysing the appropriate wave equation in curved spacetime. For example, for a scalar field $\psi(x)$ one considers
\begin{equation}\label{E:Wave}
\nabla_a\nabla^a\psi=0,
\end{equation}
where $\nabla_a$ is the covariant derivative with the Christoffel connexion associated with the spacetime metric $g_{ab}$. The probability of emission is related to the ratio of the amplitude of inward and outward radial travelling solutions to equation (\ref{E:Wave}) at spatial infinity.  It can be shown, using separation of variables, that for a non charged, spherically symmetric (Schwarzschild) black hole of mass $M$ this reduces to finding the transmission probability for equation (\ref{E:strategy}) with the Regge--Wheeler potential (see reference~\cite{Regge:1957} and,  for more recent background, reference~\cite{Boonserm:2013} and references therein):
\begin{equation}
V(r^*)=\left[1-\frac{2M}{r(r^*)}\right]\left[\frac{\ell(\ell+1)}{r(r^*)^2}+\frac{(1-s^2)2M}{r(r^*)^3}\right].
\end{equation}
Here, $r$ is the usual radial coordinate and $r^*$ the so-called tortoise coordinate. They are explicitly related by
\begin{equation}
r(r^*)=2M\left[1+ W\left(\exp\left(\frac{r^*}{2M}-1\right)\right)\right],
\end{equation}
where $W(x)$ is the Lambert $W$ function~\cite{Boonserm:2013, Lambert1, Lambert2, primes}. In addition,  $\ell$ is the principal angular momentum number, and $s$ is the spin of the particle. $s\in \{0,1,2\}$ for scalars, photons, and gravitons respectively, and $\ell\in\{s,s+1,...\}$. There are $2\ell+1$ azimuthal modes for each principle angular momentum mode, $\ell$, which in the case of spherical symmetry are equiprobable. The emission rate $\d N_s (\omega)/\d t\,\d\omega$ gives the  total probability for an emission, per unit time, per unit frequency, of a particle of spin $s$ and frequency $\omega$. It is given by the sum over the transmission probabilities $T_{\ell,s}(\omega)$, (equation (\ref{E:Tprob})), for each principal and azimuthal angular momentum mode, multiplied by the probability for a particle to be in a given mode $P_{\ell,s}(\omega)$.  That is~\cite{Hawking:1975, Hawking:1976, Hartle:1976, Page:1976a}:
\begin{equation}\label{E:NumRate}
\frac{\d N_s (\omega)}{\d t\,\d\omega}=\sum_{\ell=s}^{\infty}(2\ell+1)\;T_{\ell,s}(\omega)\;P_{\ell,s}(\omega),
\end{equation}
where for a Schwarzschild black hole and integer spin particles the probability for the particle to be in a given mode is given by the Bose--Einstein distribution,
\begin{equation}
P_{\ell,s}(\omega)=\frac{g}{2\pi}\frac{1}{\exp{(8\pi M\omega)}-1},
\end{equation}
and $g$ is the number of polarizations for a given spin $s$.
Equation (\ref{E:NumRate}) represents the rate of the emission of particles. Each particle carries one quantum of energy, $\omega$, and so the energy emission is given by
\begin{equation}\label{E:EnRate}
\frac{\d E_s (\omega)}{\d t\,\d\omega}=\omega\frac{\d N_s (\omega)}{\d t\d\omega}.
\end{equation}
Another physically important quantity is the cross--section, $\sigma(\omega)$, which represents an effective area that embodies the likelihood of a particle to be scattered, (i.e. deflected), by the black hole. This is intimately related to the probability of transmission through the potential barrier~\cite{Page:1976a}
\begin{equation}\label{E:Cross}
\sigma_s(\omega)=\pi\omega^{-2}\sum_{\ell=s}^{\infty}(2\ell+1)T_{\ell,s}(\omega).
\end{equation}
In the high frequency limit $M\omega\gg1$ this approaches the classical geometric optics cross-section $\sigma_\infty=27\pi M^2$~\cite{Page:1976a}. This can now be used to define the dimensionless cross-section, $S(x)$, by
\begin{equation}\label{E:S}
S(x)=\frac{\sigma(x)}{\sigma_\infty}=\frac{1}{27x^2}\sum_{\ell=s}^{\infty}(2\ell+1)T_{\ell,s}(x),
\end{equation}
where $x=M\omega$. Equation (\ref{E:S}) can the be used to rewrite equations (\ref{E:NumRate}) and (\ref{E:EnRate}) in dimensionless form,
\begin{equation}
M\,\frac{\d E_s (x)}{\d t\,\d\omega}=x\frac{\d N_s (x)}{\d t \,\d\omega}=\frac{g}{2\pi}\frac{27x^3S(x)}{\exp{(8\pi x)}-1}.
\end{equation}

Now the Regge--Wheeler potential is asymptotically zero at both ends, (that is, we have $V(r^*)\rightarrow0$ as $r^*\rightarrow\pm\infty$). So using equation (\ref{E:Tprob}) the transmission probabilities can be calculated from the  Bogoliubov coefficients, as given in equation (\ref{E:Transfer0}).
In this case the transfer matrix becomes
\begin{equation}\label{E:GB}
\begin{bmatrix}
\alpha & \beta^* \\
\beta & \alpha^* \\
\end{bmatrix}
=\mathcal{P}\exp\left( -\frac{i}{4x}\int_{-\infty}^{+\infty}V(u^*)
\begin{bmatrix}
1 & \exp(-4ixu^*)\\
-\exp(4ixu^*) & -1\\
\end{bmatrix}
du^* \right).
\end{equation}
where the potential,
\begin{equation}
V(u^*)=\left[1-\frac{1}{u(u^*)}\right]\left[\frac{\ell(\ell+1)}{u(u^*)^2}+\frac{(1-s^2)}{u(u^*)^3}\right],
\end{equation}
has now been re-written in terms of the dimensionless variables $u^*=r^*/2M$ and $u(u^*)=r(u^*)/2M$.
In the product calculus formalism this leads to
\begin{equation}\label{E:GB2}
\begin{bmatrix}
\alpha & \beta^* \\
\beta & \alpha^* \\
\end{bmatrix}
=\prod_{-\infty}^{+\infty}(\I +A(u^*)\;du^*),
\end{equation}
where
\begin{equation}
A(u^*)\equiv
-\frac{i}{4 x}V(u^*)
\begin{bmatrix}
1 & e^{-4ixu^*}\\
-e^{4ixu^*} & -1 \\
\end{bmatrix}.
\end{equation}
It is also possible to do a little pre-processing by changing variables in the path-ordered integral. This might somewhat help analytic insight, but does not seem to improve the numerics. See appendix \ref{A:B}.

\subsection{Numerics}

\begin{figure}[!h]
\includegraphics[width=0.325\textwidth]{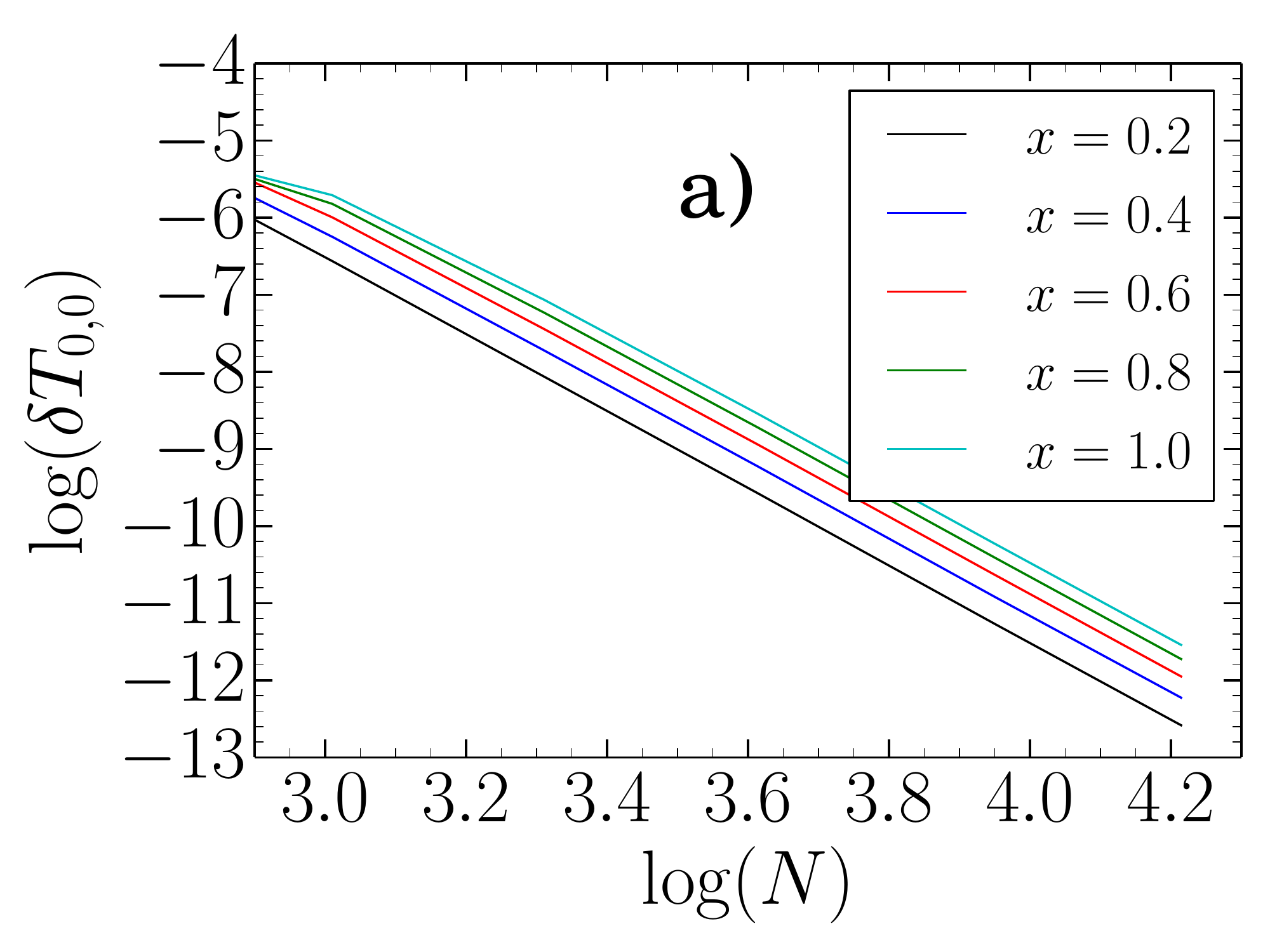}
\hfill\includegraphics[width=0.325\textwidth]{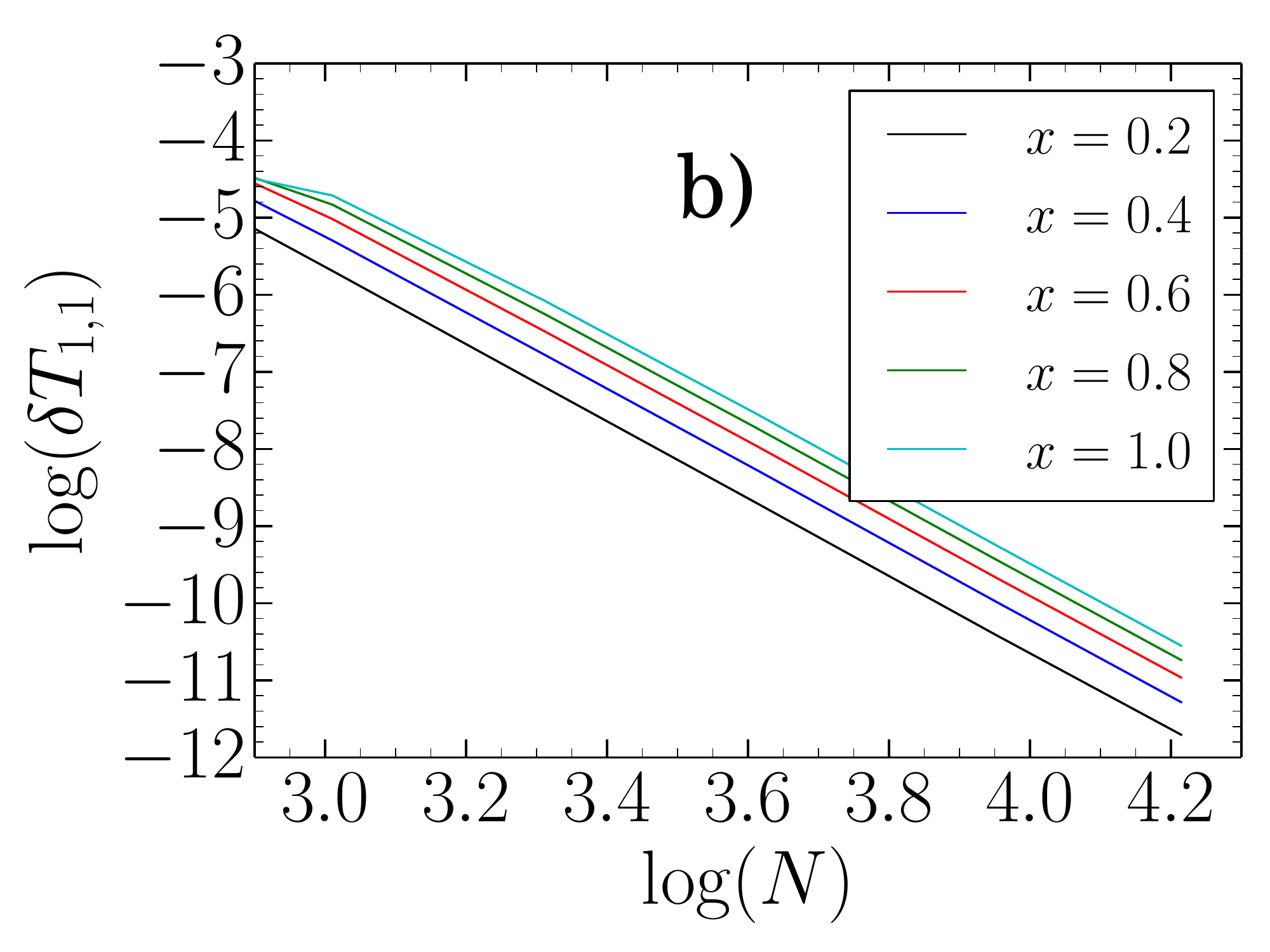}
\hfill\includegraphics[width=0.325\textwidth]{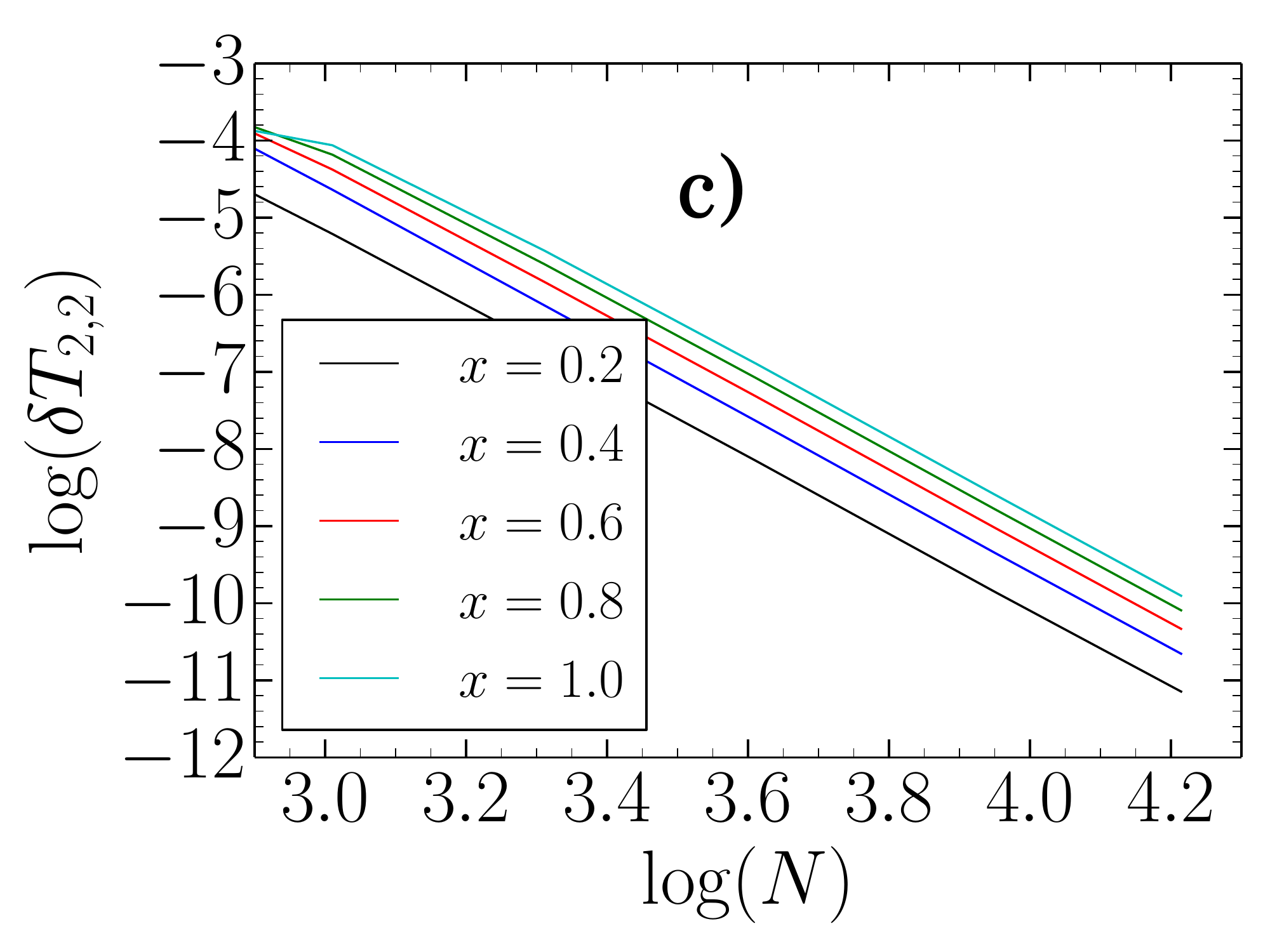}
\caption{{\bf Regge--Wheeler for Schwarzschild: Convergence tests.}\newline Convergence rates of the numerical transmission probabilities $T_{s,s}(x)$, (greybody factors),  for (a)  scalars, (b) photons, and (c) gravitons. Here we have defined  the relative error of the $N$-th approximation, (i.e. $N$ terms), as
$\delta T_{s,s}\equiv \big| T^{(N)}_{s,s}(x)-T^{(N+1)}_{s,s}(x) \big|/T^{(N)}_{s,s}(x)$.}
\label{F:Con}
\end{figure}

The calculation for the transmission probabilities for the Regge--Wheeler potential was numerically implemented in {\sf Python} by using the polynomial approximation of equation (\ref{E:Approx2}). The integration region $[-\infty,+\infty]$ was approximated by the finite range $[-50,350]$, which was found to introduce an error of at worst $\sim \O(10^{-9})$. Numerical convergence tests for the product integrals are summarized in figure \ref{F:Con}. 

\begin{figure}[!h]
\includegraphics[width=0.325\textwidth]{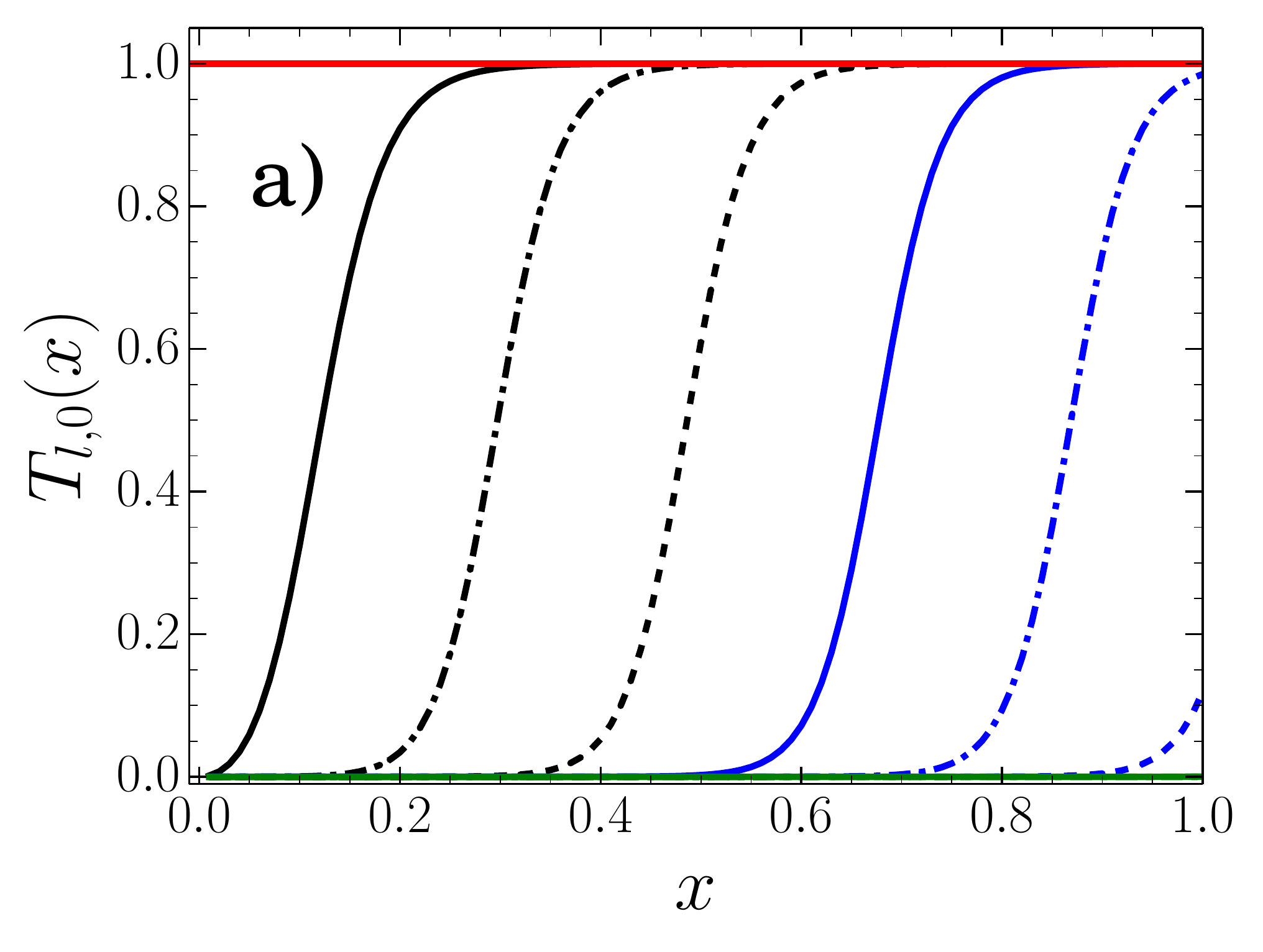}
\hfill\includegraphics[width=0.325\textwidth]{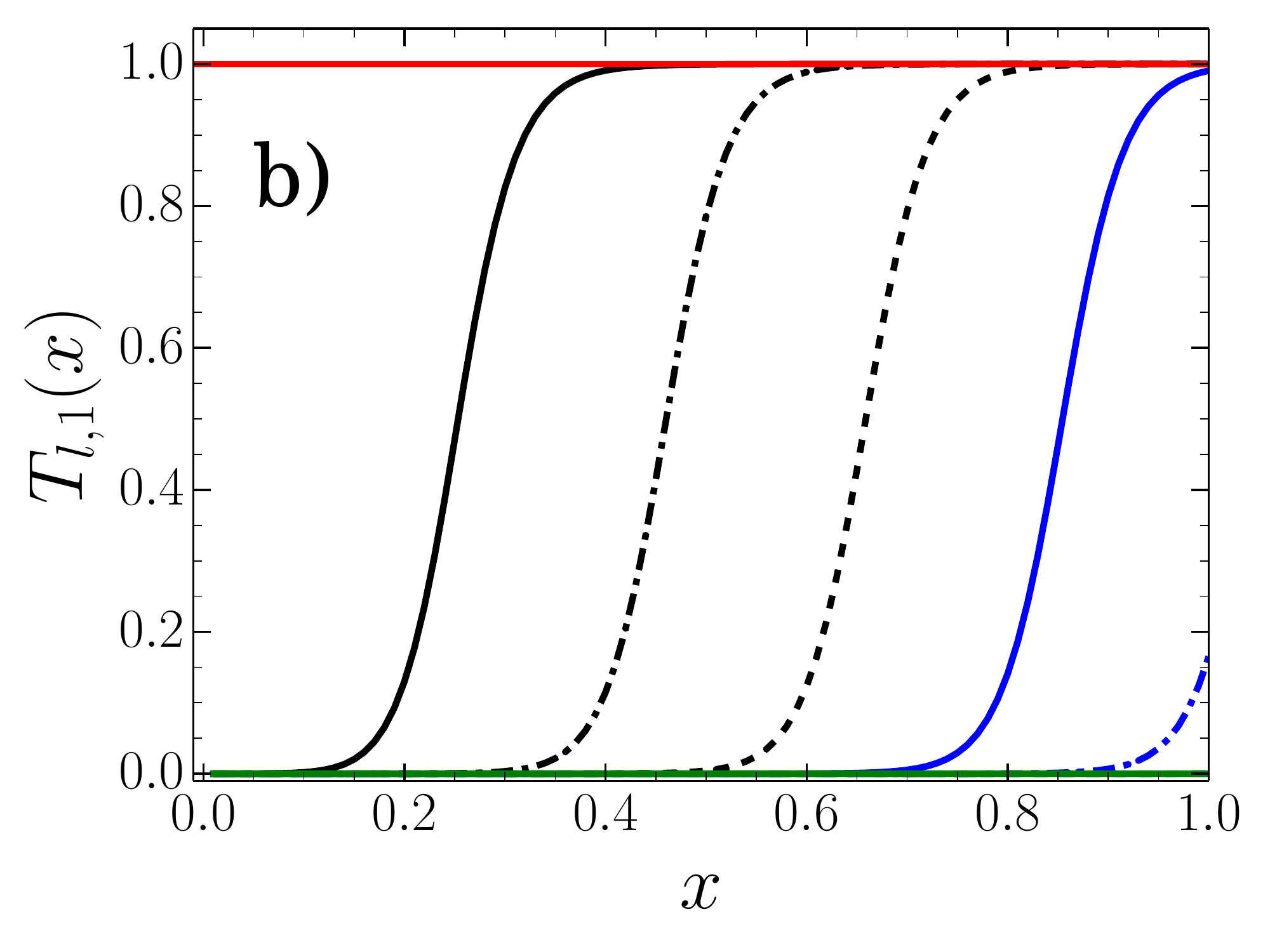}
\hfill\includegraphics[width=0.325\textwidth]{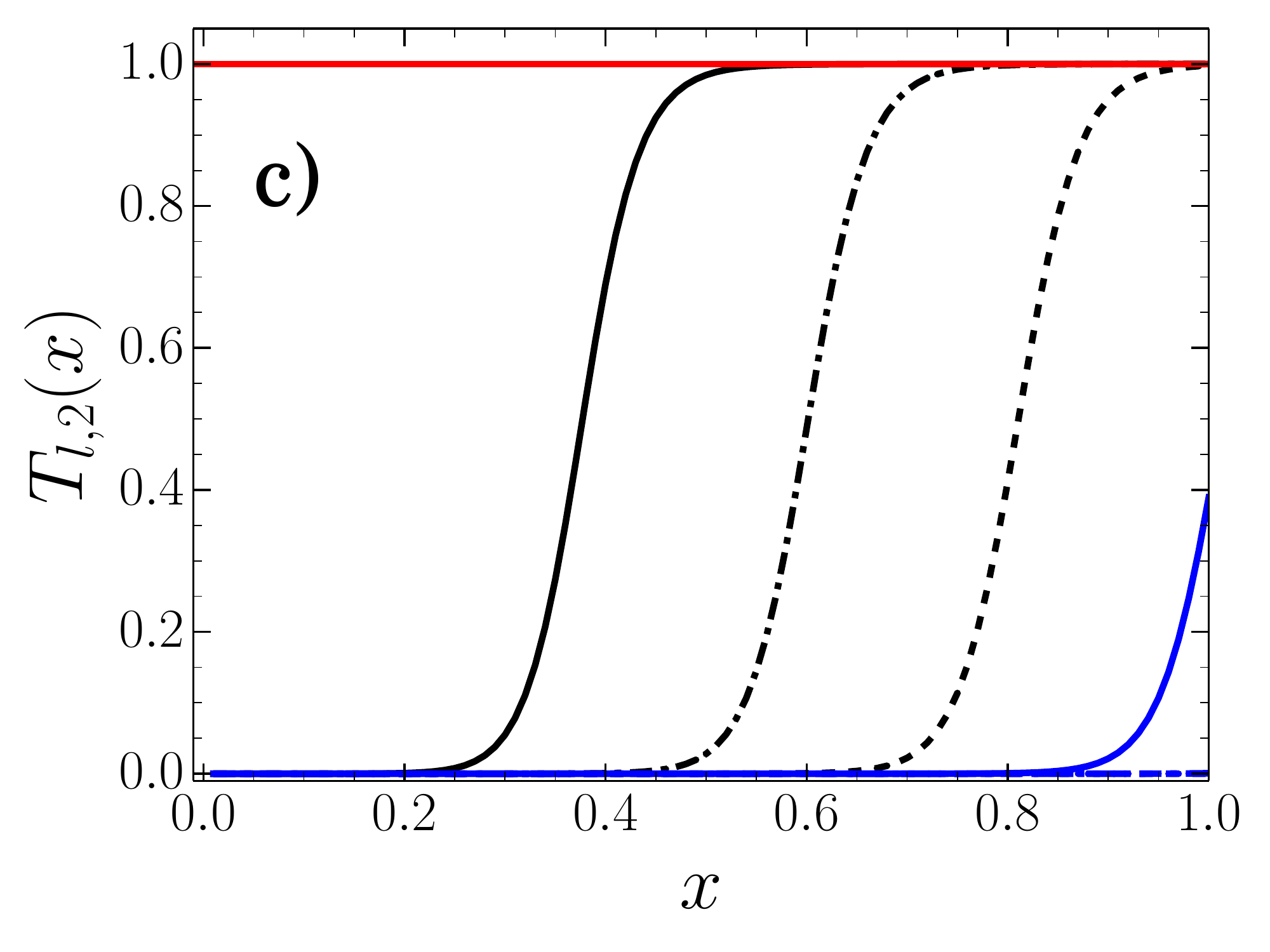}
\caption{{\bf Regge--Wheeler for Schwarzschild: Greybody factors.}\newline Plots of the transmission probabilities, $T_{\ell,s}(x)$, as a function of $x=M\omega$ for, a) scalars, \newline b) photons, and c) gravitons. The leftmost function on each plot corresponds to the $\ell=s$ transmission probability, increasing $\ell$ values occur to the right.}
\label{F:Trans}
\end{figure}

Figure \ref{F:Trans}  shows the transmission probability for each of the scalar, photon, and graviton cases. It can be seen that the $s=0$ case has transmission at the lowest frequencies, and that in each case larger $\ell$ values require higher frequencies before there is any transmission through the barrier. Furthermore  for each $\ell$ eventually (at high enough frequencies) there is complete transmission through the barrier. This can be interpreted physically by observing the potential $V(r^*)$ is lowest for the $\ell=0=s$ case and as such less energy (i.e. lower frequency) is required to pass through the barrier. For larger $\ell$ and $s$ values the potential is higher and so more energy is required. Eventually any particle species in any mode will have enough energy to completely pass through the barrier, i.e. $T_{\ell,s}(x)\rightarrow1$ as $x\rightarrow\infty$.

\begin{figure}[!h]
\begin{center}
\includegraphics[scale=0.3500]{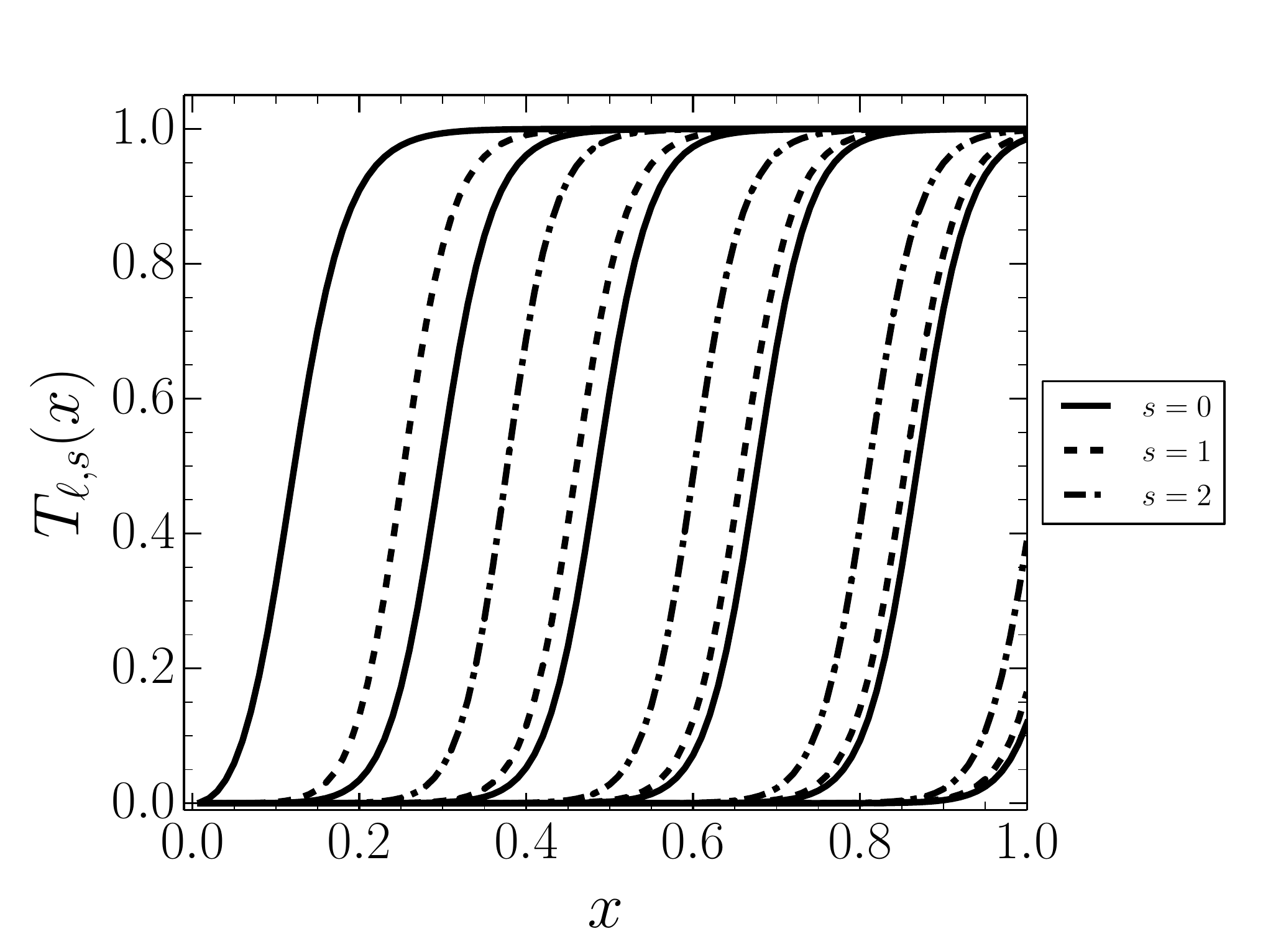}
\includegraphics[scale=0.3500]{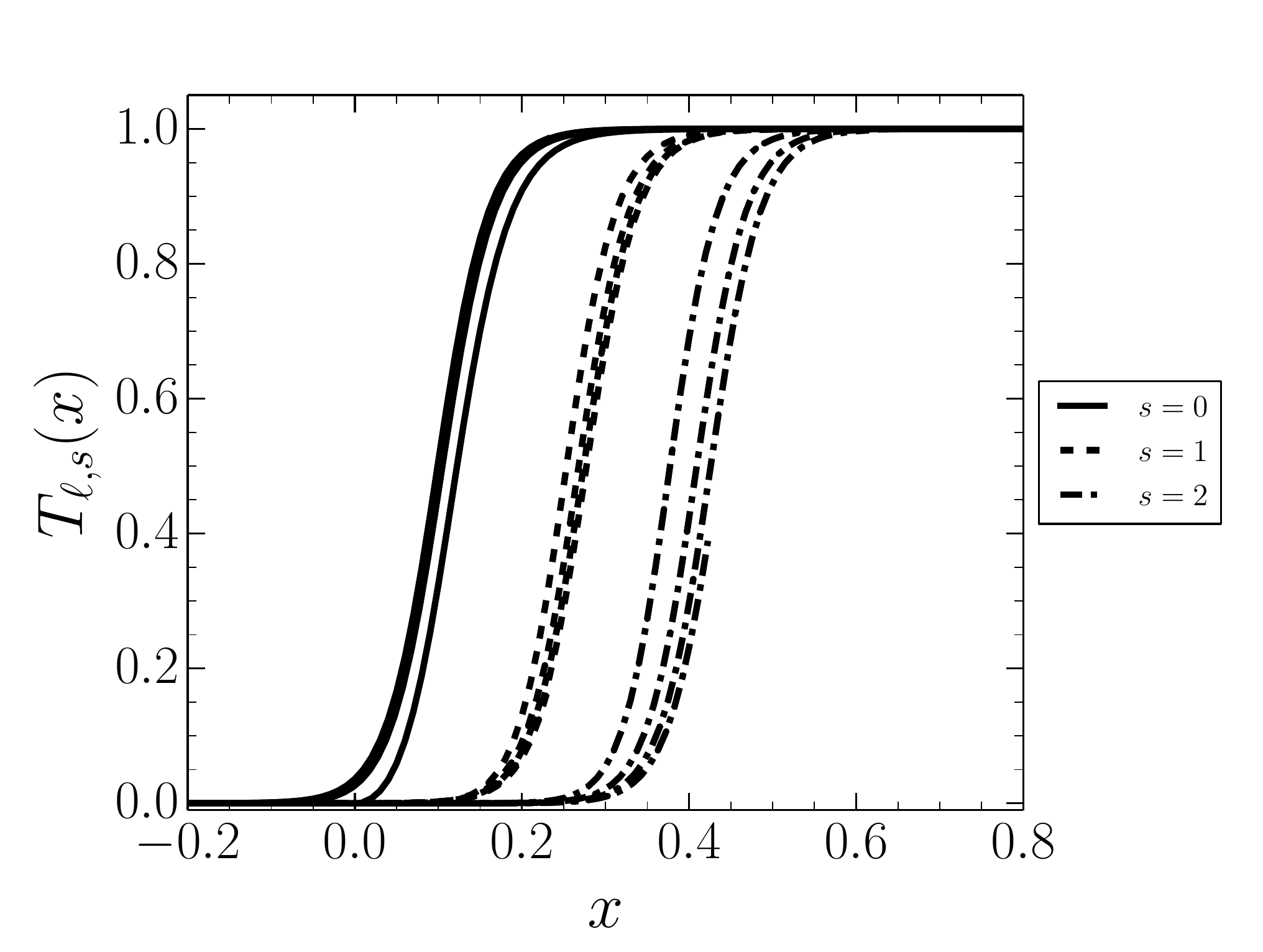}
\includegraphics[scale=0.3500]{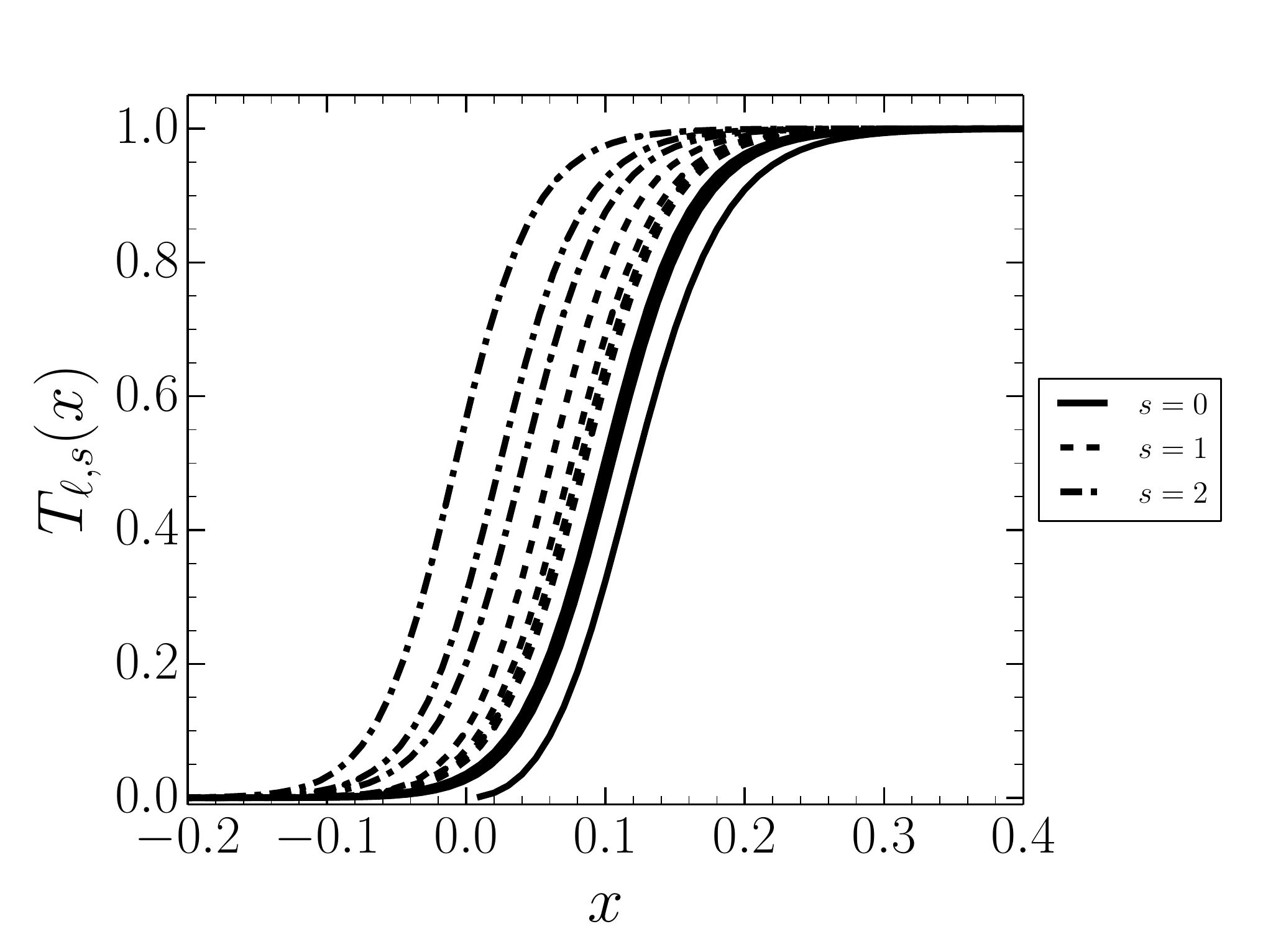}
\end{center}
\caption{\label{F:Sall}{\bf Regge--Wheeler for Schwarzschild: Superimposed greybody factors.}\newline Note the very strong similarities and relatively minor differences. The greybody factors almost seem to be translations of one another, and are very similar to suitably shifted hyperbolic tangent functions.
The left hand figure merely superimposes the greybody factors. The right hand figure translates the greybody factors to the left by $(\ell-s)/\sqrt{27}$ before superimposing them.  The bottom figure translates the greybody factors to the left by $\ell/\sqrt{27}$ before superimposing them.  
}
\end{figure}

These greybody factors (for the Regge--Wheeler potential in Schwarzschild spacetime) exhibit very strong similarities and relatively minor differences. The greybody factors almost seem to be translations of one another, and all appear to be similar to suitably shifted hyperbolic tangent functions. See figure \ref{F:Sall}. 
We shall discuss the implications of this observation more fully in the subsequent section on ``modelling''.

\begin{figure}[!h]
\includegraphics[scale=0.35]{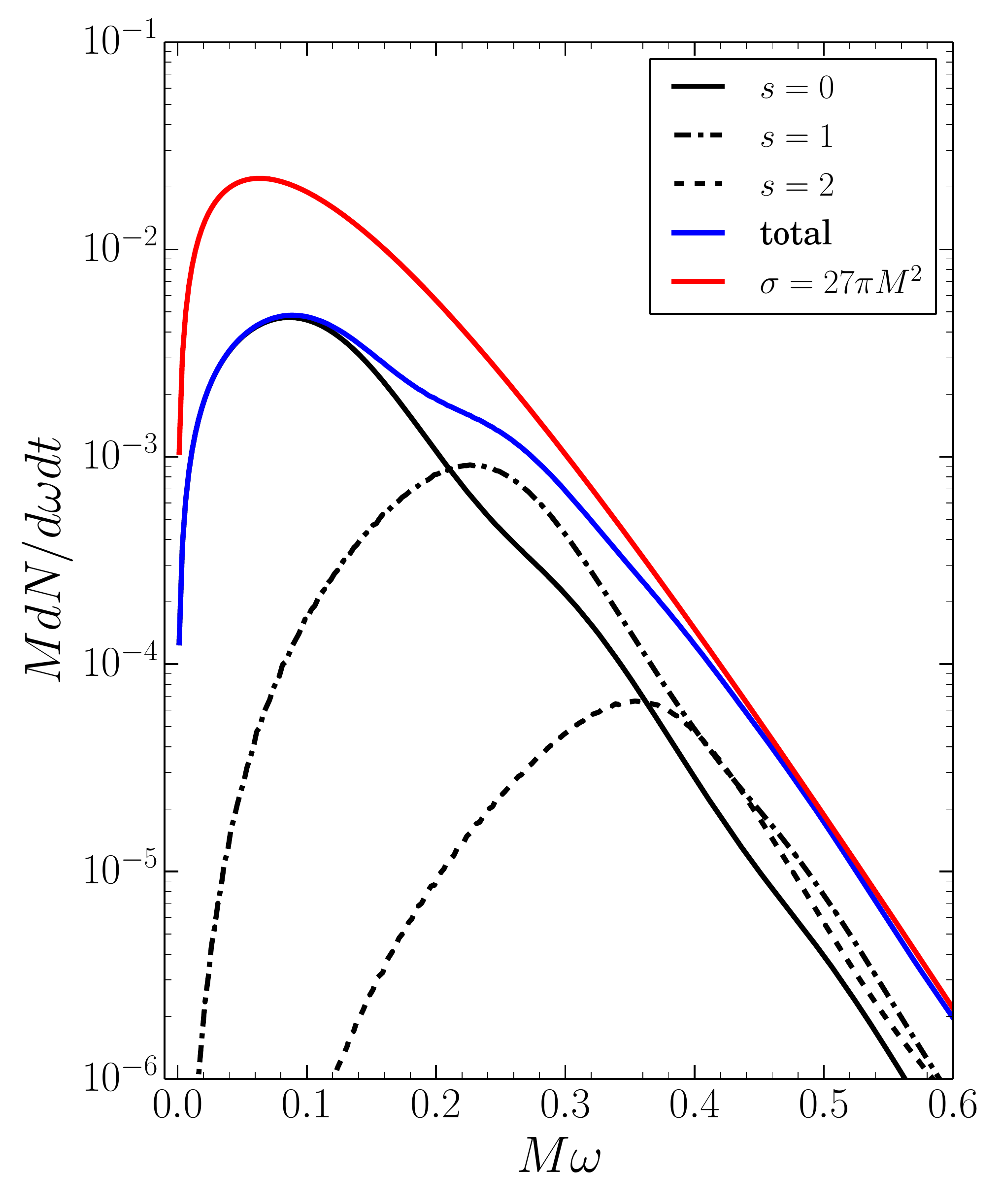}
\hfill
\includegraphics[scale=0.366]{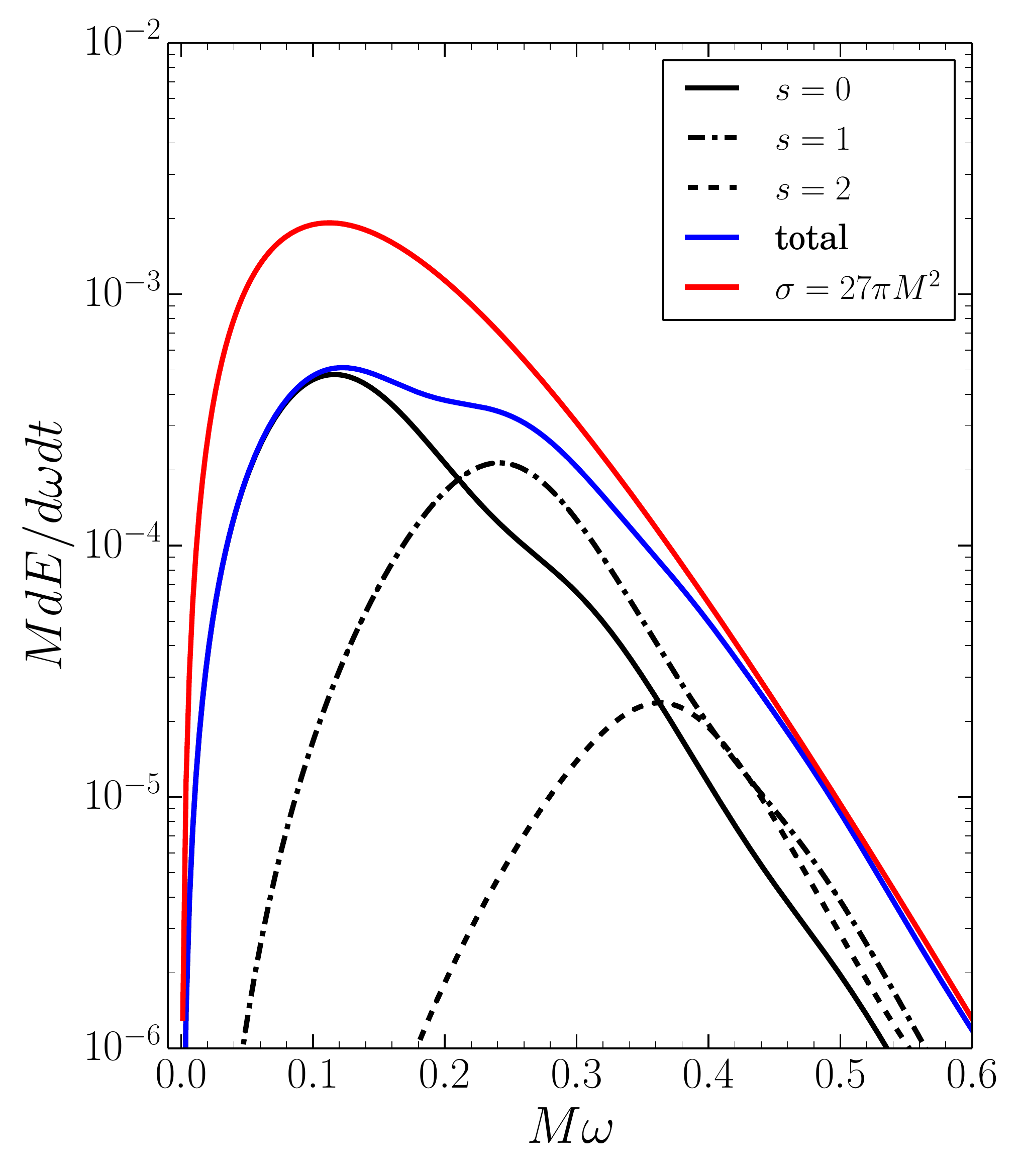}
\caption{\label{F:Num}{\bf Regge--Wheeler for Schwarzschild: Number and energy spectra.} 
\newline Plots of the number (left) and energy (right) emission spectra for a Schwarzschild black hole, see equations (\ref{E:NumRate}) and (\ref{E:EnRate}). Note the logarithmic scale. The emission spectrum is dominated by scalar particles, and emission rates decrease with increasing spin. The total emission rates (i.e. summing over all particle species)  are bounded by the sums over the geometric optics limit, $\sigma=27\pi M^2$, for each species (shown in red), and this limit is approached as $M\omega\rightarrow\infty$.}
\end{figure}

Figure \ref{F:Num} shows plots of both the number and energy emissions rates, equations (\ref{E:NumRate}) and  (\ref{E:EnRate}). It can be seen that for $x\gtrsim0.6$ the emission rate  rapidly  becomes negligible, exponentially decaying to zero. The emission of particles is dominated by scalars, and the rate reduces with increasing spin. This can be understood from figure \ref{F:Trans}, in which it can be seen that for lower spin $s$ the transmission probabilities become significant at smaller $x$; which coincides with the peak in the probability spectrum $P_{\ell,s}(x)$. For spin $s$ most of the interesting physics is concentrated in the range $x\in\left[0,{(s+1)/\sqrt{27}}\right]$, whereas for total emissivity
most of the interesting physics is concentrated in the range $x\in\left[0,{1/\sqrt{27}}\right]$.

\begin{figure}[!ht]
\centering
\includegraphics[scale=0.40]{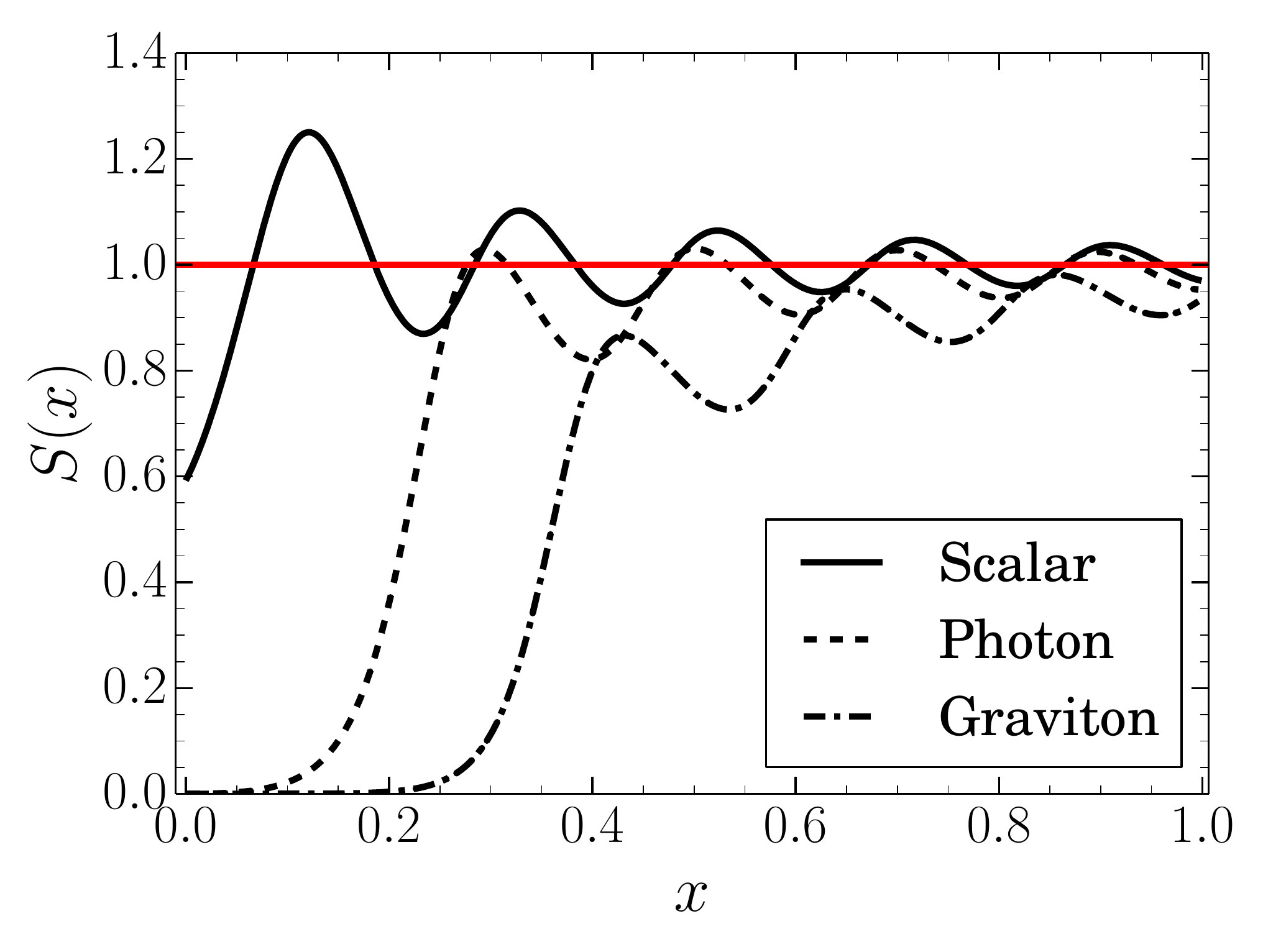}
\caption{\label{F:Scomp}{\bf Regge--Wheeler for Schwarzschild: Dimensionless cross--sections.}
\newline
The dimensionless cross--section $S(x)$, see equation (\ref{E:S}), plotted for scalars, photons, and gravitons for a Schwarzschild black hole.}
\end{figure}

Finally figure \ref{F:Scomp} shows a numerical plot of the dimensionless cross--section, for each species of particle. As $x$ increases the cross-section approaches the geometric optics limit, i.e. $S(x)\rightarrow1$. This happens more quickly for lower spins. Each species also exhibits oscillations which are due to the transmission probabilities becoming appreciable for increasing $\ell$ values, weighted by the $2\ell+1$ factor.
%

\subsection{Modelling the greybody factors and cross-section}

Given the numerical data, and in view of other information we have available, to what extent can we characterize and model salient features of the greybody factors?
In the low-frequency limit we know that~\cite{Page:1976a, Starobinsky}:
\begin{equation}
T_{\ell,s}(\omega) \approx  C_{\ell,s} \; x^{2\ell+2}; 
\qquad
\hbox{with}
\qquad
C_{\ell,s} = \left[2^{2\ell+2}(\ell-s)!\ell!(\ell+s)!\over(2\ell)!(2\ell+1)! \right]^2 .
\end{equation}

At high frequencies we know that~\cite{Sanchez:1976, Sanchez:1977a, Sanchez:1977b}: 
\begin{equation}
\label{E:Sanchez2}
T_{\ell,0}(x) \approx {1\over 1 + \exp\left\{(2\ell+1)\pi \left[1-{27 x^2\over(\ell+{1\over2})^2}\right]\right\}}; \qquad  \qquad (x \gg 1; \ell \gg 1).
\end{equation}
Unfortunately this conveys relatively little information beyond the fact that $T(x)\to1$ exponentially rapidly as $x\to\infty$. 
At intermediate frequencies rather little qualitative or quantitative information regarding the greybody factors is available. 
In contrast, for the dimensionless (scalar) cross section $S(x)$ more is known. At intermediate/high frequencies Sanchez gives the equivalent of~\cite{Sanchez:1976, Sanchez:1977a, Sanchez:1977b}:
\begin{equation}
S_{s=0}(x)  \approx 1 - \sqrt{32\over27}\; {\sin\left(2\pi\sqrt{27} x\right)\over2\pi\sqrt{27} x} ;   \qquad  x \gtrsim {1\over\sqrt{27}}.
\end{equation}
However, inspection of figure \ref{F:Num} shows that the bulk of the total emission spectrum is concentrated in the range $x\in[0,{1/\sqrt{27}}]$, where this is not all that good an approximation. In fact the Sanchez approximation predicts a negative cross section $S_{s=0}(0)<0$ at $x=0$. Can we do better?

An extremely simple toy model that captures \emph{most} (but not all) of the relevant physics is to use a simple sigmoid function and take
\begin{equation}
\label{E:Model0}
T_\mathrm{model\,0}(x) = 
{1\over \displaystyle1 + \exp\left(- w_{\ell,s} [x - x^*_{\ell,s} ]\right)} = 
{1\over2} \left[
1+ \tanh\left({w_{\ell,s}\over2}\left[x-x^*_{\ell,s} \right]\right) \right].
\end{equation}
The parameter $w_{\ell,s}$ controls the (inverse) width of the transition zone from 0 to 1, and inspection of figures \ref{F:Trans} and \ref{F:Sall} indicates that the transition zone is close to $1/\sqrt{27}$ wide for all spins and angular momenta. This corresponds to $w_{\ell,s}\approx 27$.

\begin{figure}[!h]
\centering
\includegraphics[scale=0.50]{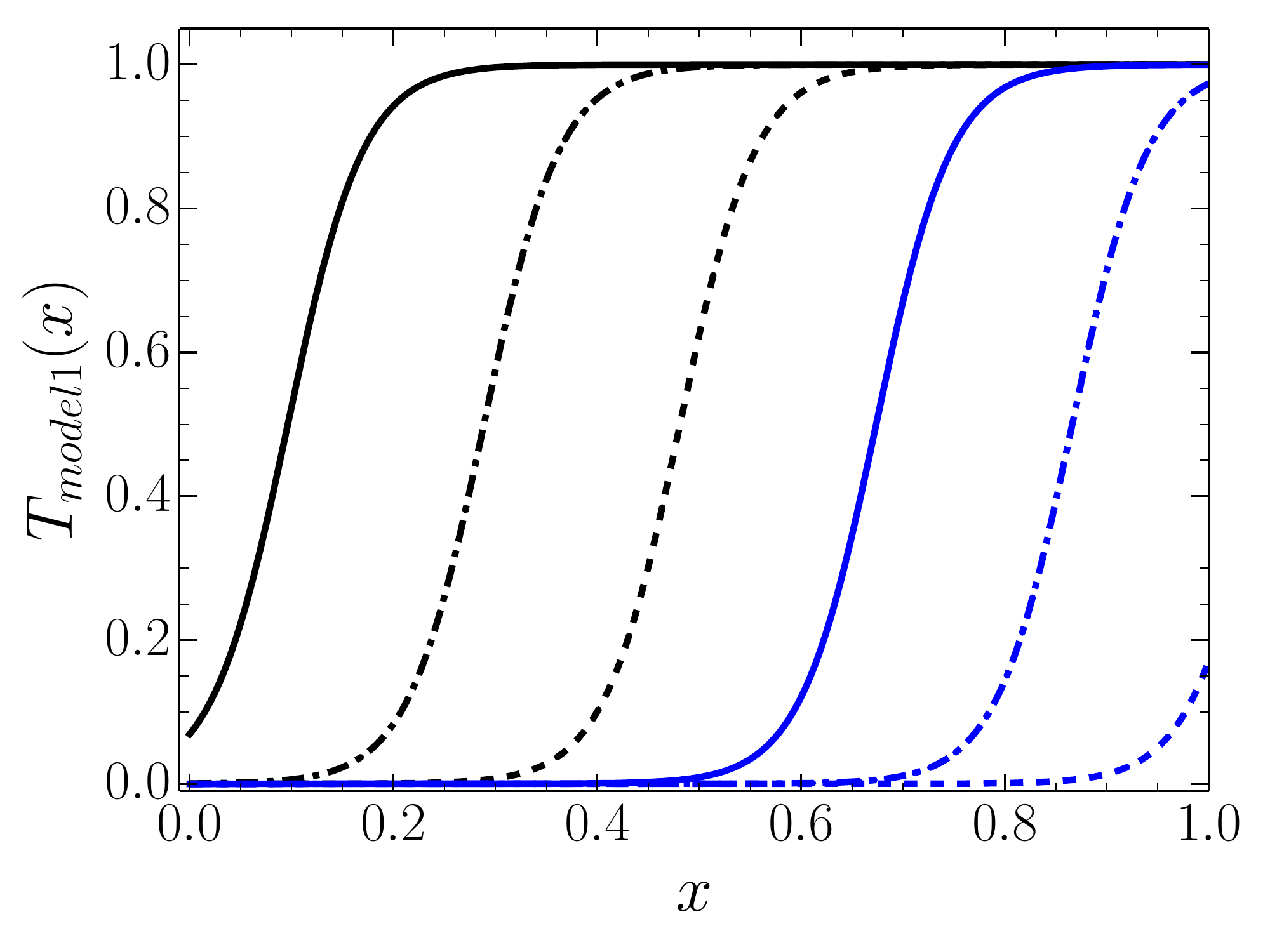}
\caption{\label{F:Model1}{\bf Regge--Wheeler for Schwarzschild: Greybody factors.}\newline
Toy model \# 1, equation (\ref{E:Model1}), for the greybody factors. Note the bad behaviour at $x=0$, though other gross features of the numerically determined greybody factors are adequately represented.}
\end{figure}

The parameter $x^*_{\ell,s}$ controls the location of the transition zone from 0 to 1.
For any $x$ only those modes with $x^*_{\ell,s}<x$ contribute appreciable to the sum in the dimensionless cross section $S(x)$. In fact a very crude approximation is 
\begin{equation}
S(x) \approx  {\sum_{\ell<\ell_\mathrm{max}(x)}  (2\ell+1) \over 27 x^2} = {\ell_\mathrm{max}(x)^2\over27x^2};
\qquad 
\ell_\mathrm{max}(x) = \max\{\ell: x^*_{\ell,s}<x\}.
\end{equation}
But $S(x)\to1$ asymptotically, so $\ell_\mathrm{max}(x)\to\sqrt{27} x$, which can be inverted to yield
\begin{equation}
x^*_{\ell,s} \to {\ell\over\sqrt{27}} \qquad \hbox{as} \qquad \ell\to\infty.
\end{equation}
This explains why the transition zones as displayed in figures \ref{F:Trans} and \ref{F:Sall} are roughly equally spaced, and explains the specific value of the spacing. In fact, inspection of figures \ref{F:Trans} and \ref{F:Sall} indicates that a tolerable approximation for all $\ell$ is
\begin{equation}
x^*_{\ell,s} \approx {\ell+ {1\over2}\over\sqrt{27}}.
\end{equation}
Thus with these specific values for the parameters our simple toy model becomes
\begin{eqnarray}
\label{E:Model1}
T_\mathrm{model\,1}(x) &=& 
{1\over \displaystyle1 + \exp\left(-27\left[x-{1\over\sqrt{27}}\left(\ell+{1\over2}\right)\right]\right)}
\nonumber\\[7pt]
&=& {1\over2} \left[
1+ \tanh\left({27\over2}\left[x-{1\over\sqrt{27}}\left(\ell+{1\over2}\right)\right]\right) \right].
\end{eqnarray}
See figure \ref{F:Model1}, and compare this simple model with figures \ref{F:Trans} and \ref{F:Sall}.

\begin{figure}[!h]
\centering
\includegraphics[scale=0.50]{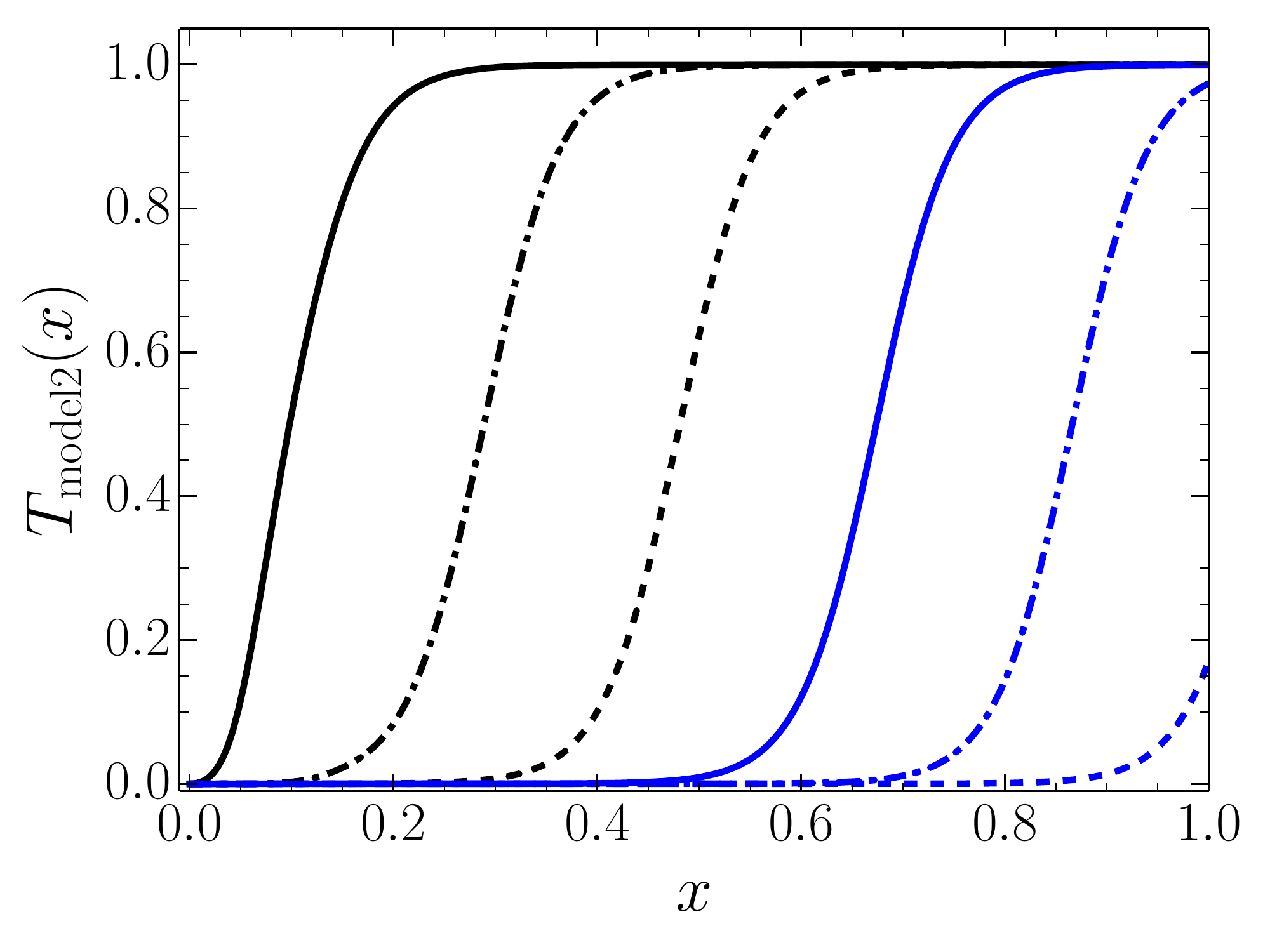}
\caption{\label{F:Model2}{\bf Regge--Wheeler for Schwarzschild: Greybody factors.}\newline
Toy model \# 2 , equation (\ref{E:Model2}), for the greybody factors. Note improved behaviour at $x=0$.}
\end{figure}

The major weakness of this simple sigmoid model is the behaviour as $x\to0$, where the non-zero limiting values of the greybody factors, $T(x\to0)\neq 0$, naively lead to an infinite cross section, $S(x\to0)\to\infty$.
The known $x\to0$ behaviour suggests we instead take something like this:
\begin{equation}
\label{E:Model2}
T_\mathrm{model\,2}(x) = 
{\tanh\left(C_{\ell,s}\; x^{2\ell+2}\left\{1 + \exp\left(\sqrt{27}\left(\ell+{1\over2}\right)\right)\right\} \right)\over 1 + \exp\left(-27\left[x-{1\over\sqrt{27}}\left(\ell+{1\over2}\right)\right]\right)}.
\end{equation}
Note this improved model simultaneously gives both the correct small $x$ behaviour, ($C_{\ell,s} x^{2\ell+2}$), the correct large $x$ behaviour ($T\to 1$), and appropriate spacing and width for the transition zones, so $S(x\to\infty)\to1$ as required. See figure \ref{F:Model2}, and compare this simple model with figures \ref{F:Trans} and \ref{F:Sall}.
The corresponding cross sections are still not so good a fit at intermediate $x$. The locations of the peaks is fine but the height of the first peak in $S(x)$ is overestimated by some 75\%.

\begin{figure}[!h]
\centering
\includegraphics[scale=0.50]{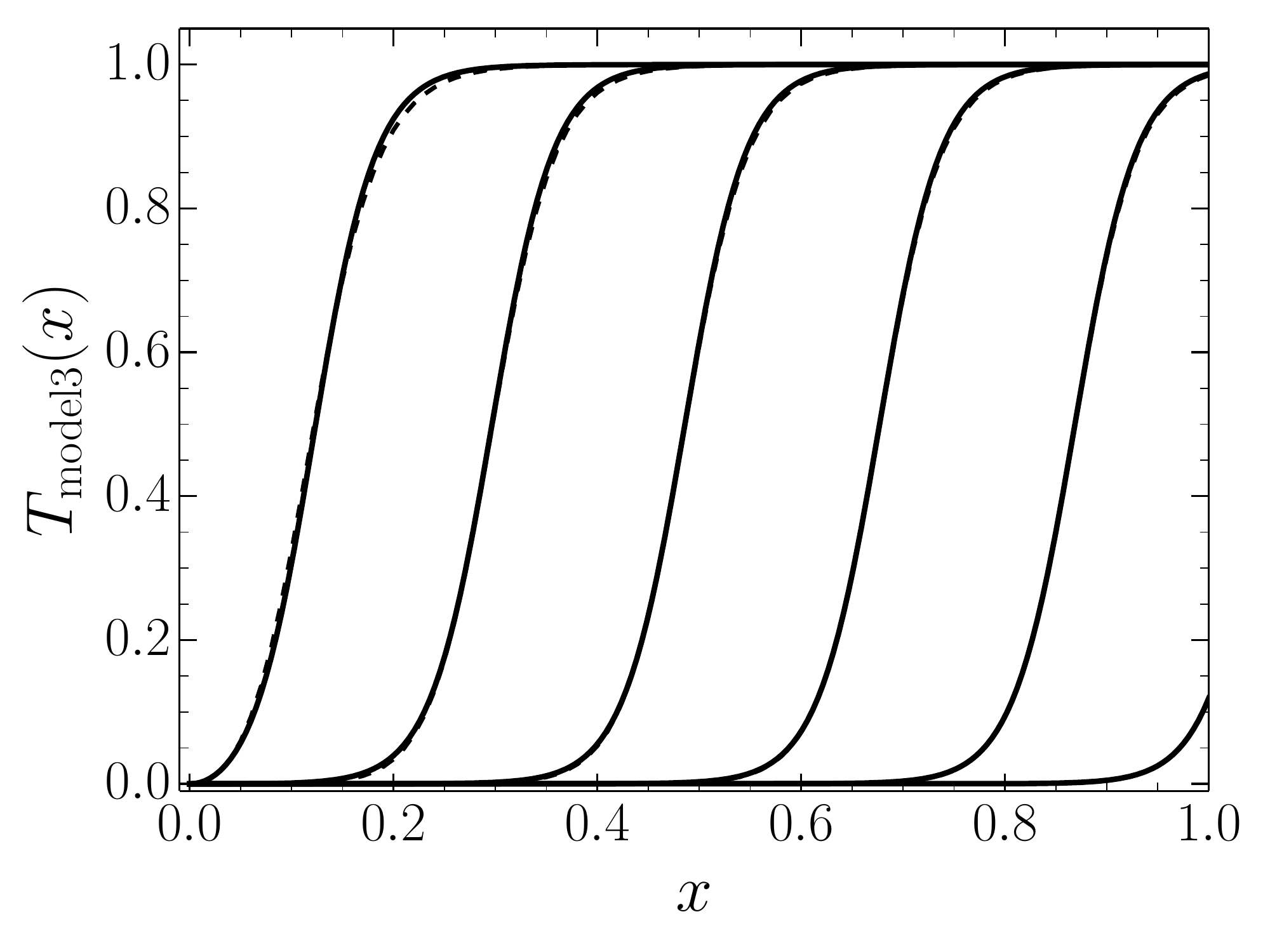}
\caption{\label{F:Model3}{\bf Regge--Wheeler for Schwarzschild: Scalar greybody factors.}\newline
Toy model \# 3, equation (\ref{E:Model3}), for the scalar greybody factors compared to the numerical data. Note the solid curves are the numerical data, while the dashed curves are from our model \# 3. The curves are often indistinguishable to the naked eye. }
\end{figure}

\begin{figure}[!h]
\centering
\includegraphics[scale=0.85]{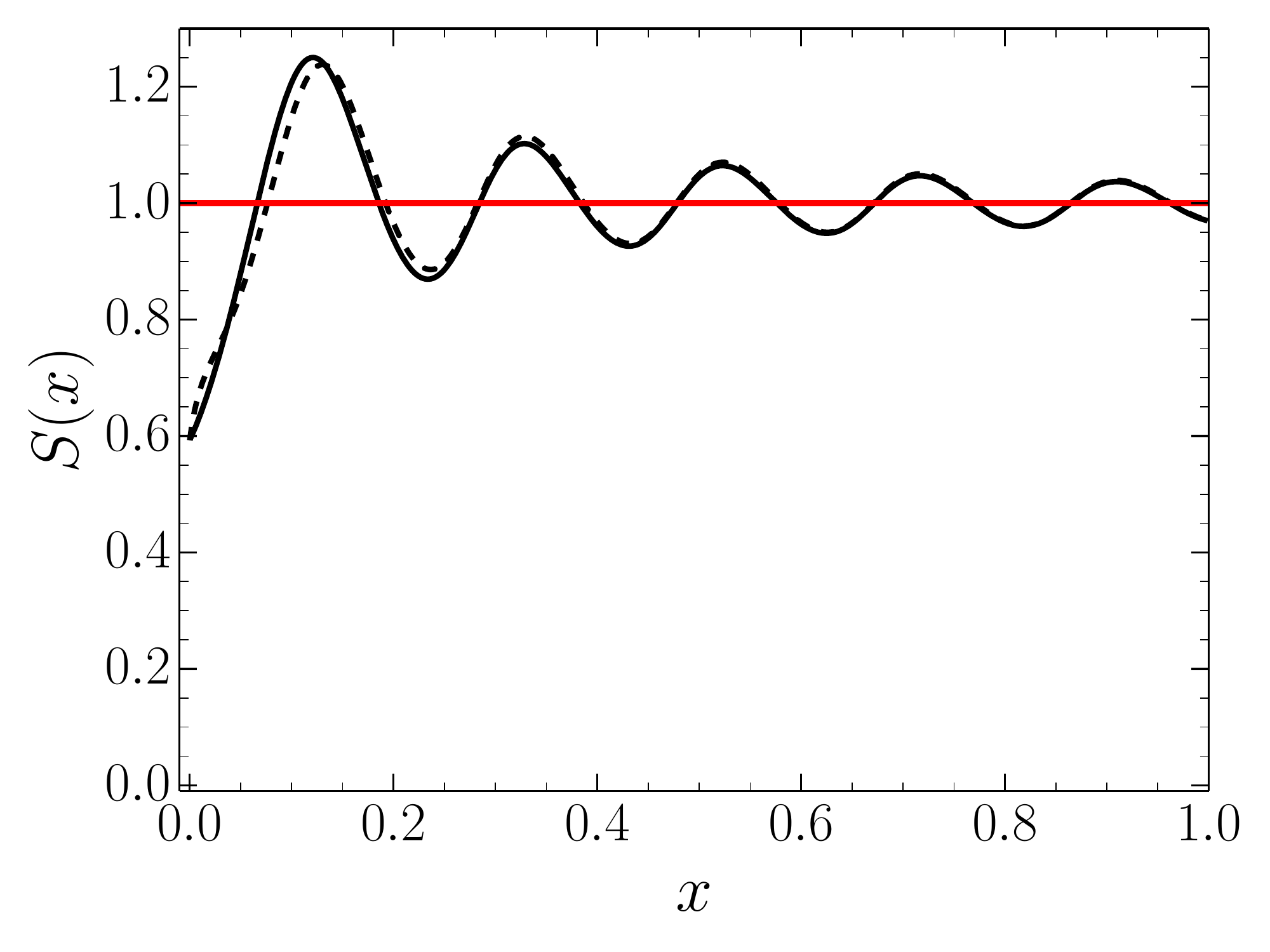}
\caption{\label{F:Model3-S}{\bf Regge--Wheeler for Schwarzschild: Scalar cross section.}\newline
Dimensionless (scalar) cross section derived from the toy model \# 3, equation (\ref{E:Model3}), for scalar greybody factors.
Note the solid curve is based on the numerical data, while the dashed curve is from our model \# 3. The fit is generally quite good and the curves are often indistinguishable to the naked eye. }
\end{figure}

In an attempt to further improve the model, one thing we looked at was the location and width of the transition zones. 
For the scalar case we found that
\begin{equation}
x^*_{\ell,s=0} \approx{1\over\sqrt{27}}  \left[ \ell+ {1\over2} + {1\over16 (\ell+{1\over2})}\right]
\end{equation}
gave a slightly better estimate for the location of the transition zones. (Presumably these are the first two terms of an asymptotic expansion.) We also found it advantageous to artificially adjust the width parameter $w_{\ell,s}\to 33$. 
Finally we tweaked the low-$x$ behaviour by noting that $[\tanh(z^{1\over n})]^n \approx \tanh z$ at low $z$, while still approaching unity at large $z$, and chose $n=3$ as a good fit. With these modifications in place we now have
\begin{equation}
\label{E:Model3}
T_\mathrm{model\,3}(x) = 
{\tanh\left[\left(C_{\ell,s}\; x^{2\ell+2}
\left\{1 + \exp\left({33\over\sqrt{27}}[\left(\ell+{1\over2}\right) +{1\over16(\ell+{1\over2})} \right)\right\}\right)^{1/3}\right]^3
\over 1 + \exp\left(-33\left[x-{1\over\sqrt{27}}\left[\left(\ell+{1\over2}\right) +{1\over16(\ell+{1\over2})}\right]   \right]\right)}.
\end{equation}
This gives a remarkably good fit to the (scalar) greybody factors and cross section. See figures \ref{F:Model3} and \ref{F:Model3-S}. While it must be admitted that the model appears complicated, there are actually only three free parameters, (the exponent $3$ we have used in the tanh, the width parameter $w_{\ell,s}\approx 33$, and the shift in the location of the switchovers). The other parameters, (the $C_{\ell,s}$, the asymptotic location of the switchovers, the presence of the number 27) are \emph{fixed} by the known asymptotic behaviour of the greybody factors and cross section. Overall, this is a quite acceptable 3 parameter fit, both to the scalar greybody factors and to the scalar cross section.

Fits to $s=1$ and $s=2$ could be developed along similar lines, (by tweaking the exponent $n$ in the tanh, the width parameter $w_{\ell,s}$, and the shift in the location of the switchovers). In the interests of brevity we restrict attention to the scalar case. 

%

\subsection{Directly modelling the cross section}

\begin{figure}[!h]
\centering
\includegraphics[scale=0.7]{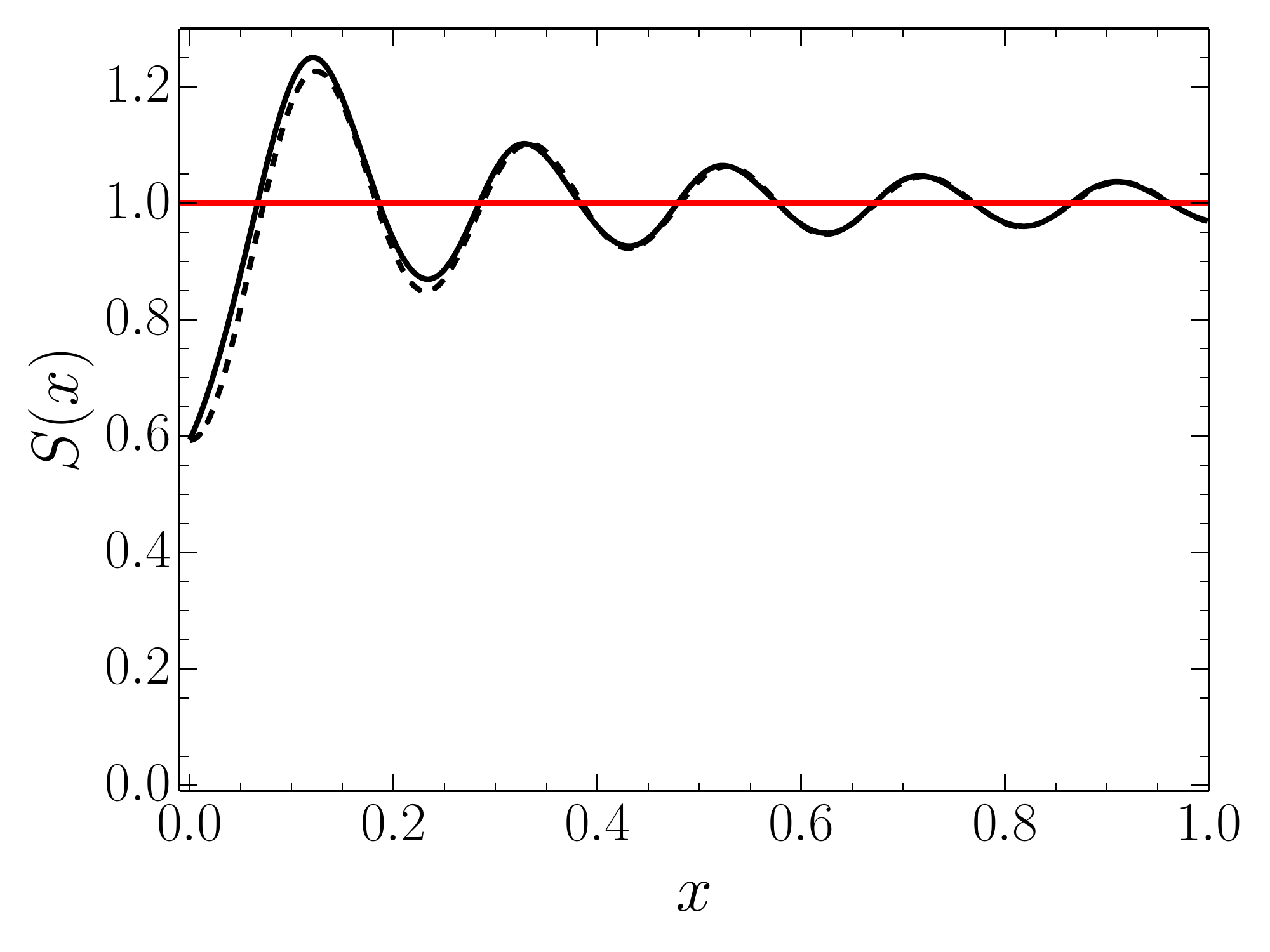}
\caption{\label{F:Modified-sanchez}{\bf Regge--Wheeler for Schwarzschild: Modified Sanchez approximation.}\newline
{The dashed curve is the model, our equation (\ref{E:sanchez-27}), the solid curve is the numerical data. The fit is generally quite good and the curves are often indistinguishable to the naked eye.}}
\end{figure}

In counterpoint, we have also looked at improving the Sanchez approximation directly (without explicitly worrying about the underlying greybody factors). The best we have been able to come up with is this:
\begin{equation}
\label{E:sanchez-27}
S(x)  \approx 1 - \sqrt{32\over27}\;{\sin\left(2\pi\sqrt{27} x\right)\over2\pi\sqrt{27} x}
+ \left\{ \sqrt{32\over27}- {11\over27} \right\}\;\exp\left(- 27x^2\right) \;{\sin\left({3\over2}\pi\sqrt{27} x\right)\over{3\over2}\pi\sqrt{27} x}.
\end{equation}
See figure \ref{F:Modified-sanchez}. The $\sqrt{32/27}$ is part of the original Sanchez approximation, and consequently the factor $\{ \sqrt{32/27}- {11/27} \}$
is not a free parameter, it is fixed by the known behaviour at $x=0$: $S(0)= 16/27$. So there are only two free parameters: The 27 in the Gaussian, and the ${3\over2}\pi\sqrt{27}$ were put in by hand, purely for \emph{observational} reasons. We know of no good analytic reason for choosing such numbers, but the outcome is impressive.
Overall, this is a quite acceptable 2 parameter fit. 

D\'ecanini, Folacci, and Rafaelli~\cite{Decanini1}, and D\'ecanini, Esposito-Farese and Folacci~\cite{Decanini2}, argue on semi-analytic grounds that the  $\sqrt{32/27}$, (which Sanchez obtains on purely numerical grounds),  should more properly be replaced by $8 \pi e^{-\pi}$. Numerically and visually these two options are indistinguishable. 
Fits to $s=1$ and $s=2$ could be developed along similar lines, (by tweaking the width and oscillations of the subdominant term). In the interests of brevity we restrict attention to the scalar case. 

\section{Discussion}

So what have we learned?
\begin{itemize}
\item Path-ordered exponentials (transfer matrices, product integrals) are an effective way of first analytically expressing the Bogoliubov coefficients associated with a scattering problem, and second can then be turned into an efficient algorithm for numerically calculating the Bogoliubov coefficients when the underlying problem is not analytically solvable. 
This observation is generic, not black-hole specific. 

\item The path-ordered exponential formalism is the only way we know of to write down a more or less explicit formula for the Bogoliubov coefficients (and hence the transmission and reflection amplitudes) associated with a scattering problem. 

\item Turning specifically to the Regge--Wheeler equation for the Schwarzschild black hole, the product integral (specifically the 5$^{th}$ order Helton--Stuckwisch algorithm, effectively a higher-order Simpson rule for product integrals), allowed us to quickly and efficiently calculate numerical greybody factors (which we then compared to older extant data from the 1970's and also used in our own recent work on the sparsity of the Hawking flux). Perhaps more interestingly, once one has enough easily manipulable data on hand, it becomes feasible to undertake some semi-analytic model building to explore the structural details of the greybody factors.

\item The 3 parameter fit we obtained to the (scalar)  greybody factors seems quite good; likewise the 
2 parameter fit we obtained to the (scalar)  cross section seems quite good.

\item  These ideas could easily be extended to Reissner--Nordstr\"om, Kerr, and Kerr--Newman black holes, and via the Teukolsky master equation to higher spins. Generic ``dirty'' black holes (black holes surrounded by matter fields) can also be dealt with as soon as one derives a suitably generalized Regge--Wheeler equation.

\end{itemize}
We hope that these observations may be of wider interest, both in a general scattering context, 
and for black hole specific calculations.

\acknowledgments

This research was supported by the Marsden Fund, 
through a grant administered by the Royal Society of New Zealand. \\
FG was also supported via a Victoria University of Wellington MSc scholarship.

\appendix
\section{Appendix: Transmission and reflection coefficients}\label{A:A}
We wish to calculate the transmission and reflection coefficients for 1D Schr\"odinger-type equations
\begin{equation}\label{E:SE}
-\frac{\d^2\psi}{\d x^2}+V(x)\,\psi(x)=E\,\psi(x),
\end{equation}
where the potential asymptotes to a constant,
\begin{equation}
\lim_{x\rightarrow\pm\infty}V(x)=V_{\pm\infty}.
\end{equation}
Later on we will assume that $V_{+\infty}=V_{-\infty}$ but in principle this is not required. In the two asymptotic regions there are two independent solutions~\cite{Visser:1999}
\begin{equation}
\psi^{\pm i}_{\pm\infty}(x)\approx \frac{\exp(\pm i\omega_{\pm\infty})}{\sqrt{\omega_{\pm\infty}}}.
\end{equation}
The $\pm i$ refers to right (left) moving modes, $e^{i\omega_{\pm\infty} x}$ ($e^{-i\omega_{\pm\infty} x}$), and $\omega_{\pm\infty}=\sqrt{E-V_{\pm\infty}}$. To analyse the transmission and reflection coefficients we consider the Jost solutions, $J_{\pm}(x)$, which are exact solutions to equation (\ref{E:SE}) satisfying
\begin{equation}
J_\pm(x\rightarrow\pm\infty)\rightarrow \frac{\exp(\pm i\omega_{\pm\infty}x)}{\sqrt{\omega_{\pm\infty}}},
\end{equation}
and
\begin{eqnarray}
J_+(x\rightarrow-\infty) &\rightarrow& \alpha\;\frac{\exp(+ i\omega_{-\infty}x)}{\sqrt{\omega_{-\infty}}}+\beta\; \frac{\exp(- i\omega_{-\infty}x)}{\sqrt{\omega_{-\infty}}}, 
\\
J_-(x\rightarrow+\infty) &\rightarrow& \alpha^*\;\frac{\exp(- i\omega_{+\infty}x)}{\sqrt{\omega_{+\infty}}}+\beta^*\; \frac{\exp(+ i\omega_{+\infty}x)}{\sqrt{\omega_{+\infty}}}.
\end{eqnarray}
Here $\alpha$ and $\beta$ are the Bogoliubov coefficients, which are related to the reflection and transmission amplitudes by
\begin{equation}\label{E:BC}
r=\frac{\beta}{\alpha};\qquad t=\frac{1}{\alpha};
\end{equation}
These Bogoliubov coefficients are for incoming/right moving waves which are partially scattered and transmitted by the potential $V(x)$.
The reflection and transmission probabilities are then given by
\begin{equation}\label{E:TR}
R=|r|^2; \qquad T=|t|^2.
\end{equation}
That is, the probability for an incident particle to be reflected off or transmitted through the potential $V(x)$ is given by $R$ or $T$ respectively.
Note that by definition the sum of the probabilities for a particle to be reflected and transmitted must be unity:
\begin{equation}\label{E:Prob}
R+T=1\quad \iff \quad |\alpha|^2-|\beta|^2=1.
\end{equation}
Now the second order Schr\"odinger equation (\ref{E:SE}) can be written as a Shabat--Zakharov system of coupled first order differential equations~\cite{Boonserm:2010}.
To do this write the wave function as
\begin{equation}
\label{E:Jost}
\psi(x)=a(x)\,\frac{\exp(+i\varphi)}{\sqrt{\varphi'}}+b(x)\,\frac{\exp(-i\varphi)}{\sqrt{\varphi'}},
\end{equation}
where $a(x)$, $b(x)$ are arbitrary functions, ``local Bogoliubov coefficients'', and the auxiliary function, $\varphi(x)$, is chosen such that it has a non zero derivative and
\begin{equation}
\varphi'(x)\rightarrow\omega_{\pm\infty}\quad \text{as} \quad x\rightarrow\pm\infty.
\end{equation}
 To reduce the number of degrees of freedom we can impose the gauge condition,
\begin{equation}\label{E:Gauge}
\frac{\d}{\d x}\left(\frac{a}{\sqrt{\varphi'}}\right) e^{+i\varphi}+\frac{\d}{\d x}\left(\frac{b}{\sqrt{\varphi'}}\right) e^{-i\varphi}=0.
\end{equation}
We now define $\omega(x)^2\equiv E-V(x)$, and $\rho\equiv \varphi''+i[\omega^2(x)-(\varphi')^2]$. Then  substitute equation (\ref{E:Jost}) into equation (\ref{E:SE}).  Using the gauge condition, equation  (\ref{E:Gauge}), one obtains the following system of equations:
\begin{equation}
\frac{\d}{\d x} 
\begin{bmatrix}
a(x) \\
b(x)\\
\end{bmatrix}
= \frac{1}{2\varphi'}
\begin{bmatrix}
i\Im[\rho] & \rho\exp(-2i\varphi)\\
\rho^*\exp(2i\varphi) & -i\Im[\rho] \\
\end{bmatrix}
\begin{bmatrix}
a(x)\\
b(x)\\
\end{bmatrix} .
\end{equation}
This has the formal solution~\cite{Visser:1999},
\begin{equation}
\begin{bmatrix}
a(x_f)\\
b(x_f)
\end{bmatrix}
=E(x_i,x_f)
\begin{bmatrix}
a(x_i)\\
b(x_i)
\end{bmatrix} ,
\end{equation}
 in terms of a generalized position-dependent transfer matrix,
\begin{equation}\label{E:Transfer}
E(x_i,x_f)=
\mathcal{P}\exp\left(\int_{x_i}^{x_f} \frac{1}{2\varphi'}
\begin{bmatrix}
i\Im[\rho] & \rho\exp(-2i\varphi)\\
\rho^*\exp(2i\varphi) & -i\Im[\rho] \\
\end{bmatrix}
\d x \right).
\end{equation}
Here ``$\mathcal{P}\exp$'' denotes a path ordered exponential operation.
In the limit $x_i\rightarrow-\infty$, $x_f\rightarrow+\infty$ this becomes an exact expression for the Bogoliubov coefficients:
\begin{equation}\label{E:Transfer2}
\begin{bmatrix}
\alpha & \beta^* \\
\beta & \alpha^* \\
\end{bmatrix}
= E(\infty,-\infty)
=\mathcal{P}\exp\left(\int_{-\infty}^{+\infty} \frac{1}{2\varphi'}
\begin{bmatrix}
i\Im[\rho] & \rho\exp(-2i\varphi)\\
\rho^*\exp(2i\varphi) & -i\Im[\rho] \\
\end{bmatrix}
\d x \right).
\end{equation}
In the case $V_{-\infty}=V_{+\infty}$ there is a natural choice for the auxiliary function, simply take $\varphi(x)\equiv\omega x$, where for simplicity we have written $\omega\equiv\omega_{\pm\infty}$. With this choice equation (\ref{E:Transfer}) can be seen to reduce to
\begin{equation}\label{E:Transfer3}
E(x_i,x_f)=
\mathcal{P}\exp\left(-\frac{i}{2\omega} \int_{x_i}^{x_f} (V(x)-V_\infty)
\begin{bmatrix}
1 & e^{-2i\omega x}\\
-e^{2i\omega x} & -1 \\
\end{bmatrix}
\d x \right),
\end{equation}
while
\begin{equation}\label{E:Transfer4}
\begin{bmatrix}
\alpha & \beta^* \\
\beta & \alpha^* \\
\end{bmatrix}
=
\mathcal{P}\exp\left(-\frac{i}{2\omega} \int_{-\infty}^{\infty} (V(x)-V_\infty)
\begin{bmatrix}
1 & e^{-2i\omega x}\\
-e^{2i\omega x} & -1 \\
\end{bmatrix}
\d x \right).
\end{equation}
Note the formalism is extremely general and flexible, the current application to greybody factors is just one example of what can be done. See for instance references~\cite{Visser:1999, Boonserm:2010, Boonserm:2013}, related formal developments in references~\cite{Boonserm:2008a, Boonserm:2008c, Boonserm:2009}, and various applications in references~\cite{Boonserm:2008b, Boonserm:2014a, Boonserm:2014b}.

\section{Appendix: Formal developments for Regge--Wheeler}\label{A:B}

We are specifically interested in the problem
\begin{equation}
\left\{ {\d^2\over\d r_*^2} + \omega^2 - V(r_*) \right\} \psi(r_*) = 0,
\end{equation}
where the Regge--Wheeler potential is 
\begin{equation}
V(r_*) =  \left\{1-{2M\over r}\right\}  \Bigg\{{\ell(\ell+1)\over r^2} + {(1-s^2) 2M\over r^3} \Bigg\},
\end{equation}
and the tortoise coordinate is
\begin{equation}
{\d r_*\over \d r } = {1\over 1-{2M\over r} }.
\end{equation}
We know
\begin{equation}
\left[ \begin{array}{cc}\alpha&\beta^*\\ \beta&\alpha^* \end{array} \right]  
= {\mathcal{P}} \exp \left\{ - {i\over2\omega} \int_{-\infty}^{+\infty} V(r_*) \left[ \begin{array}{cc}1 & e^{-2i\omega r_*}\\ -e^{2i\omega r_*} & -1 \end{array} \right]  \d r_* \right\}.
\end{equation}
Therefore, changing variables $r_*\to r$, we have:
\begin{equation}
\left[ \begin{array}{cc}\alpha&\beta^*\\ \beta&\alpha^* \end{array} \right]  
= {\mathcal{P}} \exp \left\{ - {i\over2\omega} \lint_{\!\!\!\!\!\!2M}^\infty   
\Bigg\{{\ell(\ell+1)\over r^2} + {(1-s^2) 2M\over r^3}  \Bigg\}
\; \left[ \begin{array}{cc}0 & e^{-2i\omega r_*(r)}\\ -e^{2i\omega r_*(r)} & 0 \end{array} \right]  \d r \right\}.
\end{equation}
But, writing the tortoise coordinate $r_*$ in terms of the $r$ coordinate
\begin{equation}
r_*(r) =  r +2M \ln\left({r\over2M}-1\right),
\end{equation}
so
\begin{equation}
e^{2i\omega r_*(r)} = e^{2i\omega r} \left({r\over2M}-1\right)^{4i\omega M}.
\end{equation}
Therefore, somewhat more explicitly, we have
\begin{eqnarray}
\left[ \begin{array}{cc}\alpha&\beta^*\\ \beta&\alpha^* \end{array} \right]  
&=& {\mathcal{P}} \exp \left\{ - {i\over2\omega} \lint_{\!\!\!\!\!\!2M}^\infty   
\Bigg\{{\ell(\ell+1)\over r^2} + {(1-s^2) 2M\over r^3}  \Bigg\} \;  \qquad\qquad
\right.
\nonumber\\
&&
\left.
\vphantom{\lint}
\qquad\qquad
\left[ \begin{array}{cc}1 & e^{-2i\omega r} \left({r\over2M}-1\right)^{-4i\omega M}\\ 
-e^{2i\omega r} \left({r\over2M}-1\right)^{4i\omega M} & -1 \end{array} \right]  \d r \right\}.
\end{eqnarray}
Now perform another change of variables (to make everything dimensionless)
\begin{equation}
x = \omega M; \qquad u = {r\over2M}; \qquad 2 x u = \omega r.
\end{equation}
Then
\begin{eqnarray}
\left[ \begin{array}{cc}\alpha&\beta^*\\ \beta&\alpha^* \end{array} \right]  
&=& {\mathcal{P}} \exp \left\{ - {i\over 4 x} \lint_{\!\!\!\!\!\!1}^\infty   
\Bigg\{{\ell(\ell+1)\over u^2} + {(1-s^2)\over u^3}  \Bigg\}\; 
\right.
\nonumber\\
&&
\qquad\qquad\qquad
\left. 
\vphantom{\lint}
\left[ \begin{array}{cc}1 & e^{-4ix u} \left(u-1\right)^{-4i x}\\ 
-e^{4ix u} \left(u-1\right)^{4ix} & -1 \end{array} \right]  \d u \right\}.
\end{eqnarray}
Finally set $w=1/u$ so that
\begin{eqnarray}
\left[ \begin{array}{cc}\alpha&\beta^*\\ \beta&\alpha^* \end{array} \right]  
&=& {\mathcal{P}} \exp \left\{ - {i\over 4 x} \lint_{\!\!\!\!\!\!\!0}^1   \left\{\ell(\ell+1) + (1-s^2) w  \right\} \; 
\right.
\nonumber\\
&&
\qquad\qquad
\left.
\vphantom{\lint}\left[ \begin{array}{cc}1 & e^{-4ix/w} \left(1-w\over w\right)^{-4i x}\\ 
-e^{4ix/w} \left(1-w\over w\right)^{4ix} & -1 \end{array} \right]  \d w \right\}.
\end{eqnarray}
The benefit of these transformations is that the integral now runs over a finite range; the disadvantage is the rapid oscillations near the endpoints. From a theoretician's perspective this is now probably the most explicit formulation of the problem one can realistically hope for. 


\end{document}